# Community Report From the Biosignatures Standards of Evidence Workshop

**Authors:** Victoria Meadows (University of Washington), Heather Graham (NASA-GFCS), Victor Abrahamsson (JPL NASA), Zach Adam (Arizona University), Elena Amador-French (JPL NASA/Caltech), Giada Arney (NASA-GSFC), Laurie Barge (JPL NASA), Erica Barlow (UNSW/PSU), Anamaria Berea (George Mason University), Maitrayee Bose (Arizona State University), Dina Bower (University of Maryland, College Park/NASA-GSFC), Marjorie Chan (University of Utah), Jim Cleaves (Carnegie Institution for Science), Andrea Corpolongo (University of Cincinnati), Miles Currie (University of Washington), Shawn Domagal-Goldman (NASA-GSFC), Chuanfei Dong (Princeton University), Jennifer Eigenbrode (NASA-GSFC), Allison Enright (U. New Brusnwick), Thomas J. Fauchez (NASA-GSFC), Martin Fisk (Oregon State University-Emeritus), Matthew Fricke (University of New Mexico), Yuka Fujii (National Astronomical Observatory of Japan), Andrew Gangidine (US Naval Labs), Eftal Gezer (Gebze Technical University), Daniel Glavin (NASA-GSFC), John Lee Grenfell (DLR-Germany), Sonny Harman (NASA-Ames), Roland Hatzenpichler (Montana University), Libby Hausrath, (University of Nevada, Las Vegas), Bryana Henderson (NASA JPL/Caltech), Sarah Stewart Johnson (Georgetown University), Andrea Jones (NASA-GSFC), Trinity Hamilton (University of Minnesota), Keyron Hickman-Lewis (Natural History Museum, London), Linda Jahnke (NASA-Ames), Sarah Stewart Johnson (Georgetown University), Betul Kacar (University of Arizona), Ravi Kopparapu (NASA-GSFC), Christopher Kempes (Santa Fe Institute), Adrienne Kish (Muséum National d'Histoire Naturelle), Joshua Krissansen-Totton (UC Santa Cruz), Wil Leavitt (Dartmouth), Yu Komatsu (National Astronomical Observatory of Japan), Tim Lichtenberg (University of Oxford), Melody Lindsay (Bigelow Laboratory for Ocean Sciences), Catherine Maggiori (McGill University), David Marais (NASA-Ames), Cole Mathis (Santa Fe Institute, Arizona State University), Yuki Morono (Japan Agency for Marine-Earth Science and Technology), Marc Neveu (University of Maryland/NASA-GSFC), Grace Ni (University of Maryland), Conor Nixon (NASA-GSFC), Stephanie Olson (Purdue University), Niki Parenteau (NASA-Ames), Scott Perl (JPL NASA), Richard Quinn (SETI institute), Chinmayee Raj (Georgia Institute of Technology), Laura Rodriguez (JPL NASA), Lindsay Rutter (University of Tsukuba), McCullen Sandora (University of Southern Denmark), Britney Schmidt (Cornell University), Eddie Schwieterman (UC-Riverside), Antigona Segura (Universidad Nacional Autónoma de México), Fatih Şekerci (Istanbul Technical University), Lauren Seyler (Stockton University), Harrison Smith (Earth-Life Science Institute), Georgia Soares (University of New South Wales Sydney), Sanjoy Som (NASA-Ames), Shino Suzuki (JAXA), Bonnie Teece (University of New South Wales), Jessica Weber (JPL NASA), Felisa Wolfe Simon(Independent consultant), Michael Wong (University of Washington), Hajime Yano (JAXA)

**Science Organizing Committee Co- Chairs:** Victoria Meadows (University of Washington) & Heather Graham (Catholic University of America/NASA-GSFC)

**Science Organizing Committee:** Giada Arney (NASA-GSFC), Dina Bower (NASA-GSFC) Bradley Burcar(Georgetown University), Thomas Fauchez (NASA-GSFC), Yuka Fujii (National Astronomical Observatory of Japan), Lee Grenfell (DLR-Germany), Sonny Harman (NASA Ames), Sarah Stewart Johnson (Georgetown University), Josh Krissansen-Totton (UC-Santa Cruz), Graham Lau (Blue Marble Space Institute of Science), Melody Lindsay (Bigelow Laboratory for Ocean Sciences), Grace Ni (University of Maryland), Stephanie Olson (Purdue), Niki Parenteau (NASA – Ames), Heike Rauer (DLR-Germany), Britney Schmidt (Cornell University), Eddie Schwieterman (UC Riverside), Lauren Seyler (Stockton University), Amy Smith (Woods Hole Oceanographic Institution), Andrew Steele (Carnegie Institution for Science), Sara Walker (Arizona State University), Mike Wong (University of Washington)

**Standards of Evidence Workshop participants (in addition to the coauthors):** Daniel Angerhausen (Eidgenössische Technische Hochschule (ETH) Zürich), Ricardo Arevalo Jr (University of Maryland), Lorenzo Aureli (University of Tuscia), Max Bernstein (NASA), Julie Bevilacqua (Georgetown University), Linda Billings (National Institute of Aerospace), Tim Brooke (Public Health England), Christina Buffo (Georgia Institute of Technology), Roger Buick (University of Washington), Bradley Burcar (Georgetown University), William Brinckerhoff (NASA-GSFC), Sherry Cady (Portland State University), Ilaria Catanzaro (University of Tuscia) , Barbara Cavalazzi (University of Bologna), Sounak Chakraborty (Public Health Engineering), Nipun Chandrasiri (University of Peradeniya), Aditya Chopra (Australian National University), Charles Cockell (University of Edinburgh), David Des Marais (NASA-Ames), Ilankushali Elavarasan (Space Development Nexus), Gabrielle Engelmann-Suissa (University of Washington), Caroline Fontana (University of Arizona), Katherine Freeman (Penn State University), Caroline Freissinet (CNRS), Kosuke Fujishima (Earth Life Science Institute, Japan), Lisseth Gavilan (NASA- AMES), Aaron Gronstal (NASA-Ames), Samantha Gilbert (University of Washington), Jackie Goordial (University of Guelph), James Green (NASA), Melissa Guzman(LATMOS), Ashley Hannah (University of Maryland), Fariha Hasan (Quaid-i-Azam University), Lindsay Hays (NASA), Ozge Kahraman Ilikkan (Baskent University), Hiroyuki Ishikawa (NAO-Japan), Estelle Janin (University College London), Devanshu Jha (MVJ College of Engineering, Bangalore, India), Stephen Kane (UC Riverside), Marc Kaufman (Freelance Journalist), Christopher Kempes (Santa Fe Institute), Melissa Kirven-Brooks (NASA-Ames), Jon Lima-Zaloumis (Arizona State University), Ralph Lorenz (Johns Hopkins Applied Physics Lab). Gordon Love (UC Riverside), Paul Mahaffy (GFCS), Richard Mathies (UC Berkeley), Taro Matsuo (Nagoya University), Abel Mendez (UPR Arecibo), Keyron Molaverdikhani (LSW Heidelberg), Mariam Naseem (Blue Marble Space Institute of Science), Stevanus K. Nugroho (NAO/National Institutes Of Natural Sciences Astrobiology Center), Donald Obenhuber (US FDA), Courtney O'Connor (JPL NASA/Caltech), Karen Olsson-Francis (The Open University), Masahi Omiya (NAO-Japan), Alicia Ostrowska (Chalmers University), Dhruvik Pampaniya (Vyavsayi Vidhya Pratishthan Engineering College), Victor Parro (Centro de Astrobiologia, CSIC-INTA), Joey Pasterski (University of Illinois at Chicago), Shuvam Paul (Calcutta University), Alexander Pavlov (NASA-GSFC), Taylor Plattner (Georgia Institute of Technology), Natalie Punt (Fidum Veterinary), Sascha Quanz (ETH Zurich), Heike Rauer (DLR Insitute of Plaentary Research, Berlin), Tyler Robinson (Northern Arizona University), Sarah Rugheimer (University of Oxford), Daniella Scalice (NASA), Susanne Schwenzer (The Open University), Ismael Acosta Servetto (Blue Marble Space Institute of Science), Amrita Singh (Farrell Day, LLC), Jan Spacek (Firebird Biomolecular Sciences), Eva Stueeken (University of St. Andrews), Jill Tarter (SETI), Elizabeth Tasker (JAXA/ISAS), Yuichiro Ueno (Tokyo Institute of Technology), Pilar Vergeli (Arizona State University), Mary Voytek (NASA), Frances Westall (CNRS), Loren Williams (Georgia Institute of Technology), Alexandra Witze (independent science journalist)

**Community Co-signers:** Redyan Ahmed (Indian Institute of Science Education and Research), Vladimir Airapetian (NASA-GSFC), Ariel Aptekmann (Rutgers University), Mauro Barbier (Universidad de Atacama, Chile), Jason Barnes (University of Idaho), Andre Belem (Fluminense Federal University), Rosalba Bonaccorsi (NASA-AMES), Eric Schuyler Borges (Northern Arizona University), Eric Boyd (Montana State University), Bob Bruner (Denver Museum of Nature and Science), Fabiana Canini (University of Tuscia), Brook Carruthers (The University of Arizona), Mark Elowitz (NASA), Humberto Carvajal (Universidad Simón Bolívar), Alexandre Champagne (University of Montreal), David De la Cruz (Universidad Nacional de San Luis), David Dubois (NASA-AMES), Jamie Elsila (NASA-GFSC), Jeff Errington (Newcastle University), Alberto Fairen (Cornell University), David Fike (Washington University in St Louis), Sian Ford (McMaster University), Erin Frates (Boston University), Doug Galante (Brazilian Energy Center), Hector Gilberto Vazquez-Lopez (University of Washington), Laura Garcia-Gomez (University of Málaga), Felipe Gomez (Centro de Astrobiología), Ed Goolish (NASA), Natalie Grefenstette (Santa Fe Institute), John Hamilton (University of Hawaii at Hilo), Michael Himes (University of Central Florida), Julie Huber (Woods Hole Oceanographic Institution), Hiroshi Imanaka (SETI), Akos Kereszturi (Research Centre for Astronomy and Earth Sciences), Nancy Kiang (NASA- GISS), Serena Kim (University of Arizona), Ludmilla Kolokolova (University of Maryland), Kostas Kostantinids (Georgia Institute of Technology), Owen Lehmer (NASA-Ames), Briley Lewis (UCLA), Tim Livengood (University of Maryland/NASA-GSFC), Mercedes Lopez (Harvard-Smithsonian Center for Astrophysics), Justin Lawrence (Georgia Institute of Technology), Jacob Lustig-Yaeger (Johns Hopkins University Applied Physics Laboratory), Irma Lozado-Chavez (University of Leipzig) , Michael Malaska (JPL NASA), Jeff Marlow (Boston University), Muammar Mansor (University of Tuebingen), Franck Marchis (SETI), Jean-Luc Margot (UCLA), Jesus Martinez-Frias (Universidad Complutense de Madrid), Eva Mateo-Marti (Centro de Astrobiologia), Kathleen McIntyre (University of Central Florida), Ian Miller (Carina Chemical Laboratories), Dante Minniti (Universidad Andres Bello), Valeska Molina (Universidad de Antofagasta, Gijs Mulders (Universidad Adolfo Ibañez), Aaron Noell (JPL NASA), Patricia Nunez (Universidad Nacional Autónoma de México), George Profitiliotis (National Technical University of Athens) , Kerry Ramirez (GeoControl Systems), Chris Reinhard (Georgia Institute of Technology), Daniella Reis (Cornell University), David Renshaw (Royal Astronomical Society), Petra Rettberg (DLR-Germany), Amy Riches (University of Edinburgh/SETI), Anna Sage Ross-Browning (University of Idaho), Surangkhana Rukdee (Max Planck Institute for Extraterrestrial Physics), John Rummel (Friday Harbor Partners LLC), Mauro Ferreira Santos (JPL NASA), Ricardo Santos (University of Lisbon), George Schaible (Montana State University), Peter Schroedl (Boston University), Pritam Kumar Sett (NASA), Sunanda Sharma (JPL NASA), Svetlana Shkolyar (University of Maryland, College Park/NASA-GSFC), Alexandre Simionovici (University of Grenoble-Alpes), Caitlyn Singam (University of Maryland), John Spear (Colorado School of Mines), Brent Stockwell (Columbia University), Alexis Templeton (University of Colorado), Joseph Trigo-Rodriguez (Institute of Space Sciences), Luca Tonietti (University of Milan-Bicocca), Gareth Trubl (Lawrence Livermore National Laboratory), Ilhuiyolitzin Villicana-Pedraza (New Mexico State University), Daniel Viudez-Moreiras (Centro de Astrobiología), Vishnu Viswanathan (University of Maryland Baltimore County/NASA-GSFC), Sawsan Wehbi (University of Arizona), Aileen Wee (Blue Marble Space Institute of Science), James Welsh (Loyola University Chicago), James Windsor (Northern Arizona University), Robin Wordsworth (Harvard University), Shuhai Xiao (Virginia Tech), David Yenerall (Georgia State University)

"No single effect, experiment, or paper provides definitive evidence about its claims. Innovation identifies possibilities. Verification interrogates credibility. Progress depends on both."

### Executive Summary

The search for life beyond the Earth is the overarching goal of the NASA Astrobiology Program, and it underpins the science of missions that explore the environments of Solar System planets and exoplanets. However, the detection of extraterrestrial life, in our Solar System and beyond, is sufficiently challenging that it is likely that multiple measurements and approaches, spanning disciplines and missions, will be needed to make a convincing claim. Life detection will therefore not be an instantaneous process, and it is unlikely to be unambiguous---yet it is a high-stakes scientific achievement that will garner an enormous amount of public interest. Current and upcoming research efforts and missions aimed at detecting past and extant life could be supported by a consensus framework to plan for, assess and discuss life detection claims (c.f. Green et al., 2021). Such a framework could help increase the robustness of biosignature detection and interpretation, and improve communication with the scientific community and the public.

This task is also a timely one, as missions and missions concepts capable of making life detection measurements, and scientific claims of possible discovery of extraterrestrial life, are now increasing. Given the importance of life detection to NASA, the scientific challenges and complexity of life detection, and the difficulty in conveying this complexity to others, it is clear that the astrobiology community needs to develop guidelines for the scientific assessment of claims of biosignature detection, and to work towards developing a clear reporting protocol (See Section 1 for a more detailed motivation).

In response to this need, and the call to the community to develop a confidence scale for standards of evidence for biosignature detection (Green et al., 2021), a community-organized workshop was held on July 19-22, 2021. The meeting was designed in a fully virtual "flipped" format. Preparatory materials including readings, instructional videos and activities were made available prior to the workshop, allowing the workshop schedule to be fully dedicated to active community discussion and prompted writing sessions. To maximize global interaction, the discussion components of the workshop were held during business hours in three different time zones, Asia/Pacific, European and US, with daily information hand-off between group organizers.

Highlighting its importance and timeliness, the workshop had strong community support, with over 300 scientists across a broad range of disciplines applying to be members of the discussion. Globally, 82 applicants were selected by the co-chairs and SOC, based on proposed contribution to the workshop and programmatic considerations, which included disciplinary balance and career stage. Along with 23 SOC members and 20 invited participants, a total of 125 community members were discussion attendees in the workshop, while the remaining applicants—many of whom stated a desire to learn and observe, rather than

engage—were given the opportunity to observe and interact asynchronously, with access to supporting materials, live and recorded plenary sessions and the workshop Slack space. Attendees used a series of five 60-90 minute breakout discussions to explore issues related to the development and implementation of a generalized framework for biosignature detection, and started on a discussion of considerations for a biosignature reporting protocol.   Extensive notes were taken by all breakout groups (a total of ~379 pages), and summary presentations were drafted to share in plenary after each breakout.  During the workshop, attendees also formed writing groups to start working on aspects of the workshop report.  Even though writing the report was an optional activity, 74 members of the workshop discussion group agreed to continue weekly discussion meetings and writing sessions for two months after the workshop (until September 23, 2021).  This was effectively a series of six community focus groups working to further development of key aspects of the workshop.  A compiled and edited version of the workshop report was made available to the community for comment and co-signing on October 28, 2021.  This comment period ended on November 18.  Comments were addressed and a final version of the workshop report will be delivered to NASA HQ in late January (See the Preface for more information on logistics and attendee statistics).

**A generalized framework for biosignature detection**

The main outcome of the workshop was the development, discussion, clarification and adoption of a generalized framework for biosignature assessment, which is discussed in more detail in Section 2 of the workshop report.   The community framework has similarities with, but is not the same as that described in Green et al., (2021).   The core of the framework is a series of 5 scientific questions that guide the evaluation of a claim of life detection, which are arranged in two levels corresponding to biosignature detection (Level1: Q1,Q2) and interpretation (Level 2: Q3,Q4,Q5). The top-level questions were kept as simple as possible to aid in communicating key issues with the public, but scientific nuance was added in the question descriptions to help scientists from different fields navigate the purpose of each question.  The questions and their plain wording summaries are given below.

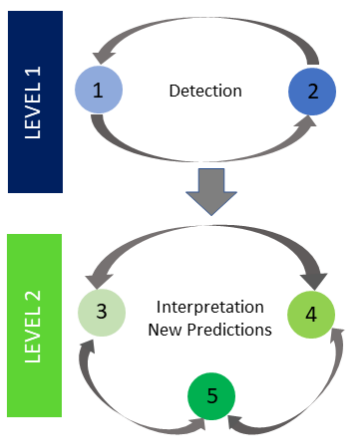

**Question 1: Have you detected an authentic signal?**
*Have you authenticated your signal, and is it statistically significant? Have you ruled out artifacts from the measurement, pre-processing and/or analysis process that might mimic a real signal?*

**Question 2: Have you adequately identified the signal?**
*Have you adequately ruled out other potential sources for this signal? For example, have you ruled out contamination in the environment, or other real phenomena that could produce a similar signal?*

 **Question 3: Are there abiotic sources for your detection?**
*Is it likely that there is a current or past environmental process, other than life, that could be producing this signal? Have you ruled out these potential false positives for the biosignature?*

**Question 4: Is it likely that life would produce this expression in this environment?**

*Given what we know about the likely environment that an organism is operating in, or would have operated in, does it make physical and chemical sense that life would produce this potential biosignature?*

**Question 5: Are there independent lines of evidence to support a biological (or non-biological) explanation?**
*Are there other measurements that provide additional evidence, or allow you to predict and execute follow-on experiments, that will help discriminate between the life or non-life hypotheses?*

While the framework Questions are presented in order, the workshop consensus was that the application of the steps do not need to be in order, and in some cases may be difficult to implement linearly. Within the Levels, measurements and processes may be cyclical or iterative, with measurements addressing several questions at once.  Measurements may also reduce, rather than enhance, the degree of certainty in a given interpretation. However, to minimize the impact on the broader community from a biosignature detection claim, it was considered desirable to have high certainty in Level 1 (the biosignature candidate detection and identification) prior to significant efforts on Level 2, where community efforts or even missions spanning a larger number of disciplines would likely be needed to evaluate whether an abiotic or biological source is the more likely explanation for the potential biosignature.

The framework asks general scientific questions that provide guidelines, or best practices, for life detection claims applicable to a variety of biosignatures and measurement techniques.  It is therefore not proscriptive of the field-specific standards for a given biosignature or measurement technique, which would need to be developed further for a given application. Nonetheless, to show proof of concept for just such an exercise, the participants were asked to test and "stress" the framework by developing worked examples of the generalized framework for well-known biosignatures in a number of different fields. Several of these worked examples are given in the workshop report.  These examples do not endorse a particular biosignature, but instead demonstrate the broad applicability of the framework. The report includes worked examples for early Earth stromatolites (Section 3.1), organics on early Earth and Mars (Section 3.2), atmospheric $O_2$ on exoplanets (Section 3.3), and agnostic *in situ* biosignatures (enantiomeric excess and isotopic composition) for a range of environments (Section 3.4). The framework also helped to guide workshop discussions on issues and guidelines for biosignature reporting, although further community work is still needed on this topic.  Indeed, the workshop participants identified a future workshop with involvement of other key stakeholders from the journalism and publishing communities as a high priority follow-on activity.

Although the biosignature framework does not provide specific standards of evidence for a given measurement, the workshop discussions coalesced around some broad guidelines for rigorous biosignature assessment.  The importance of portraying the life detection process as a multi-step, multi-mission quest was widely recognized and endorsed as being good for our science, as well as allowing clearer communication with the public about life detection milestones.  Question 1 emphasizes the need to obtain a biosignature detection that is recognized by the relevant field to be of high significance, and the importance of independent verification of the discovery.  Question 2 elevates the importance of field-relevant practices for reducing contamination, ensuring accurate characterization and recording of the context of a sample.  It also calls out the value of multiple additional, independent measurements to help confirm the biosignature, and/or rule out the potential for environmental contamination or misidentification. In addressing Question 3, best practices includes an extensive search for abiotic synthesis processes, and understanding the stability of a biosignature under the conditions in which is found. Since a high level of interdisciplinarity is needed to most accurately interpret the biosignature in the context of its environment, best practices include collaborating with a broad team of experts, and particularly those highly knowledgeable in the biosignature's environment. Question 4 emphasizes that the life hypothesis needs to be tested in its own right as a possible explanation, and not simply used as the hypothesis of last resort after other known processes have been ruled out. Best practices include habitability assessment of the current or past environment in which the biosignature is generated, and a working model for biosignature production that is physically and chemically plausible with the known environment. Finally, in Question 5, the importance of developing testable hypotheses, and the key role of multiple lines of evidence in enhancing biosignature credibility is emphasized. Best practices include but are not limited to making data/methods available, and supporting and encouraging independent confirmation of key findings.

The participants discussed how the framework for biosignature assessment could have multiple applications, including vetting and reporting biosignature discoveries and developments, and also potentially guiding different stages of mission development, from concept to operation. Being able to clearly articulate where the current status of a claim of life detection would fall in the assessment framework would help to more accurately communicate the significance of, and our confidence in, the discovery. This framework can be used irrespective of whether there is a need to communicate an initial claim of discovery, or a more thoroughly vetted result with independent confirmation, and/or ongoing interpretation. The framework could also be used to develop a mission concept's top-level science goals and instrument suite, and the specific investigations which could meet those goals. It may also help guide and ensure the inclusion of measurement capabilities for verification, contamination identification, and environmental characterization, which would be needed to confirm biosignature detection, and help with biosignature interpretation (see Section 4 for a more detailed discussion).

**Towards a biosignature reporting protocol**
The workshop provided a first opportunity to explore potential elements for best practices for an improved biosignature reporting protocol, but it was agreed that a the development of a cohesive, community endorsed protocol will need further discussion amongst the astrobiology community, as well as with other key stakeholders, such as journalists and publishers. The workshop attendees discussed the potential for miscommunication when communicating with both other scientists and the general public about the nature and certainty of any dataset or interpretation supporting a life detection claim. Workshop participants discussed potential mechanisms to help support the verification and interpretation of results, as well as ideal goals for reporting to the scientific community as well as to the general public. Participants identified obstacles to more effective collaboration and reporting as well as steps that could incentivize the community to work towards ideal verification and reporting goals. The great importance of collaboration in verification was a strong theme, and section writers researched similar attempts in other fields to develop community standards, reporting "best practices", and collaboration programs (Section 5). Key mechanisms that the group identified include funding to support verification and assessment science, encouragement to voluntarily share pre-publication data and results for verification purposes, and modifications to publication structures that could accommodate the publication of concurrent discovery and verification papers. Participants also suggested that community-led scientific groups and professional societies could establish and maintain organizational structures to help support those wishing to engage in a verification process. Scientists could also partner with journals to provide specialized subject matter experts for peer review.

**Further Work**

After this initial discussion, several important topics were identified for further development in subsequent workshops or other fora. Three of the highest priorities include further work to discuss a reporting protocol that more thoroughly addresses public communication as well as academic publications, and that includes the input and participation of stakeholders in those fields, including key members of the media and scientific publishing; to develop more detailed "worked examples" for specific biosignatures and measurement techniques; and to incorporate statistical methodology into methods to address each Question in the Framework. A brief primer of statistical methods for life detection was outlined (Section 4), but further work is needed to develop a more quantitative assessment scale (see Section 6).  Further comment from the broader community also highlighted the importance of working to include relevant technosignatures in the biosignature assessment framework, and the importance of analog studies, and further developing a theory of life.

***Preface: Workshop development, logistics and demographics***

This workshop is the result of a collaborative effort between the Nexus for Exoplanet System Science (NExSS) and the Network for Life Detection (NfoLD) Research Coordination networks. The office of the NASA Chief Scientist called for this workshop in January 2021, to encourage the science and technology communities to have a dialog about standards for communicating findings that may be interpreted as evidence of life.  An informational meeting for with the invited Co-Leads Prof. Victoria Meadows (University of Washington/NExSS) and Dr. Heather Graham (Goddard Space Flight Center/NfoLD) and others who had expressed interest was held on Feb 9, 2021 and led by Tori Hoehler (NASA Ames/NfoLD).  The workshop co-leads then convened a Science Organizing Committee and wrote a Topical Workshops, Symposia and Conference (TWSC) proposal to obtain NASA funding to support workshop logistics through the independent contractor, Know Innovation.   The Science Organizing Committee included 23 scientists with a track record of work in biosignature detection and assessment science from the early Earth, planetary science and exoplanet communities.  The SOC also included international representatives from the European and Asia/Pacific regions.

**Table 1: The Biosignatures Standards of Evidence Workshop Science Organizing Committee**

| Name | Affiliation | Expertise |
|---|---|---|
| Victoria Meadows (co-chair) | UW | exoplanets/early Earth/remote-sensing |
| Heather Graham (co-chair) | GSFC | early Earth/agnostic/in situ |
| Giada Arney | GSFC | early Earth/exoplanets/remote-sensing |
| Dina Bower | GSFC | early Earth/in situ |
| Bradley Burcar | Georgetown U. | early Earth/agnostic |
| Thomas Fauchez | GSFC | exoplanet habitability/spectral detectability |
| Yuka Fujii | NAOJ-Japan | exoplanet habitability/detectability |
| Lee Grenfell | DLR-Germany | exoplanet biosignatures/detectability |
| Sonny Harman | NASA-Ames | exoplanet biosignatures/false positives |
| Sarah Johnson | Georgetown U. | agnostic biosignatures/in situ |
| Josh Krissansen-Totton | UC-Santa Cruz | exoplanets/early Earth/biosignatures |
| Graham Lau | BMSIS | in situ/microbiology |
| Melody Lindsay | Bigelow | in situ/microbiology |

| Grace Ni | UMD | in situ/instrumentation |
| Stephanie Olson | Purdue | early Earth/exoplanet/false negatives |
| Niki Parenteau | NASA - Ames | microbiology/in situ/remote biosignatures |
| Heike Rauer | DLR-Germany | exoplanet biosignatures/detectability |
| Britney Schmidt | Cornell | in situ biosignatures/icy worlds |
| Eddie Schwieterman | UC-Riverside | novel exoplanet biosignatures/false positives |
| Lauren Seyler | Stockton U. | in situ/microbiology/icy worlds |
| Amy Smith | Bard College | in situ/microbiology |
| Andrew Steele | CIW | in situ/Early Earth/agnostic |
| Sara Walker | ASU | agnostic biosignatures, probabilistic framework |
| Mike Wong | UW | Solar System biosignature false positives |

The SOC first met on March 1, 2021. At the meeting, Meadows and Graham presented the goals of the workshop, which included the community development of a set of standards for communicating life detection discoveries, and that encompassed both a standards of evidence for biosignature assessment, and a community reporting protocol. The biosignature assessment framework was envisioned as framing the search for life as a continuum of objectives that help us best convey what we know, what we don't know and what we are going to find out, at each step of the process. Meadows and Graham also presented a strawman initial framework for discussion, comprising 5 key questions to be asked and answered in response to a claim of biosignature detection, and an outline for desired key sections of a workshop report. SOC members selected roles in sub-teams that focused on developing a list of workshop invitees, pre-meeting materials, a meeting run crew, and an archival crew. The workshop timing was discussed with the SOC and scheduled for July 19-22, 2021. This resulted in an unfortunately aggressive workshop development schedule, but was considered necessary to avoid the peak of the summer and polar field seasons, especially post-covid-vaccine, and to enhance the likelihood of European participation.

**Workshop Format and Schedule**
The workshop format, created by the co-Chairs and discussed with the SOC, was designed to be a fully virtual "flipped" format, with preparatory materials including readings, instructional videos and activities made available prior to the workshop, and with the majority of the workshop time used for active community discussion and writing. The workshop itself was supported by Know Innovation using linked Zoom breakout rooms, google docs for breakout notes and joint preparation of report-out presentations and a dedicated online platform to accommodate all of these resources. To maximize global attendance without asking attendees to disrupt their work and personal life, the discussion components of the workshop were held asynchronously in three different time zones, Asia/Pacific, European and US. The schedule of the workshop was designed so that a daily hand-off of information could be transferred between the time zone discussion sessions, usually via a debrief and/or written report between the workshop co-leads and the satellite chairs. The US session spanned 4 days, from Monday July 19 to Thursday July 22[nd], and comprised three five hour discussion sessions on days 1, 2 and 4, and a dedicated "rest day/writing day" on the Day 3. The discussion days were structured into two breakout discussions of 60-90 mins each, in which discussion notes and summary slide prepared, followed by plenary report out on the summary slides. The satellite sessions, with smaller numbers of participants, had less report out in plenary from the discussions, and therefore could also use fewer or shorter breaks, and were able to hold their sessions for 3 hrs on three consecutive days each, staggered to interleave with the US sessions, but still be within normal business hours in their respective time zones.

**Selecting Workshop Attendees**
The SOC discussed and invited 20 high priority participants to ensure that a broad range of science was covered, and that representatives from the journalism community were able to attend.   We then sent out a general call for applicants to join the workshop using a google form that was widely advertised on the Astrobiology, NfoLD, NExSS, LPI websites and Planetary Exploration Newsletter.  We received 355 applications for 80 participants slots.  The workshop attendees were capped at near 120 participants to support discussion groups of no more than 10, with enough time to report out and share results from each group on 5 separate topics in the 13 hrs of workshop discussions.   We selected 82 workshop discussion participants on their stated contribution and their work's relevance to the workshop topic, while also considering balance in field of specialty, measurement technique and career stage.   The majority of the remaining applicants had expressed an interest in learning from the workshop, rather than participating in it, and these applicants were offered observer/informatory participation in the workshop.  This was supported via YouTube LiveStream access to the discussions in real time and to recordings of those discussions, and access to the workshop Slack space for commenting/interacting with other workshop participants.  A total of 95 participants attended the US discussion session (co-chairs, SOC, invitees, accepted applicants), and 161 members of the community logged into the Slack space.

For the Asia/Pacific and European satellite sessions chairs of those sessions were identified---Yuka Fujii NAOJ/Harrison Smith (ELSI)/Hajime Yano (JAXA) and Lee Grenfell/Heike Rauer (DLR)---and these chairs invited/solicited 15 participants from each of their respective time zones for the satellite discussions.

**Table 2: Participant demographics**

125 discussion participants (95 US, 15 Asia/Pac, 15 Euro)
      19% SOC
      16% Invited
      65% Community Applicants
215 asynchronous/observing participants (40 of those signed up for Slack).

Discussion participant disciplines:
      25% exoplanet/observers
      57% solar system/planetary science
      18% early Earth/paleobiology
Discussant demographics:
      50% pre-tenure/50% post-tenure
      10% grad students, 20% post-doc, 20% early career, 20% mid-career, 30% senior.
Male/Female parity

**Pre-workshop Materials**
All participants were provided with an extensive reading list covering both remote-sensing and in situ  biosignature background/review papers, key policy documents or mission concept proposals (including the NAS exoplanet science strategy and astrobiology strategy).  Texts on challenges dealing with misinformation in science, statistical approaches to biosignature detection, and the Rio 2.0 scale proposed for assessing SETI detections were also recommended.   The reading list also included a subset of key papers on previous examples of biosignature detection claims and community response to those claims, including organics on Mars and the ALH840001 Martian meteorite, and phosphine on Venus.

Members of the community were also invited to record 15-30 minute videos that reviewed key aspects of biosignature science with the intent that this would help inform those either new to the field, or expert in a different part of the field.

**Table 3: Pre-meeting videos**

| | |
|---|---|
| Introduction and Charge | Meadows (UW) and Graham (GSFC/CUA) |
| Current status of in situ biosignature science | DesMarais (NASA Ames) |
| Current status of remote-sensing biosignature science | Domagal-Goldman (GSFC) |
| History of the Viking Biology Experiments | Quinn (NASA Ames) |
| The Controversy and Legacy of ALH840001 | Steele (Carnegie Science) |
| Phosphine on Venus | Meadows (UW) |
| SETI Frameworks for Technosignature Assessment | Tarter (SETI) |
| Statistics, Theory and Life Detection | Kempes (Santa Fe Institute) |
| The Rise of Misinformation In & About Science | West (UW) |
| Microbial Safety and Quality Science for the Unknown | Brooks (Public Health England) |

The videos garnered ~875 views as of November 2021

In addition to the readings and videos, the SOC held a "writing practice session" on June 28[th] where attendees practiced essentially the first discussion session from the workshop and started writing material that could later be used in the workshop community report. Approximately 20 people attended this practice session.

**Workshop Scope and Content**
Due to the extremely limited time available for this virtual workshop (13 hrs of interaction time) the workshop was necessarily focused on biosignature assessment and reporting protocols. This narrowed focus excluded discussions of how definitions of life and the origin of life impacted biosignature assessment, discussions or promotion of optimum or novel biosignatures to be searched for, and it did not include technosignatures in the assessment framework development. However, many of these topics could be considered for future workshops to broaden the topics discussed and stakeholder communities, and previous work on technosignature assessment scales were included as critical background material and context for the biosignature discussions.

The core of the workshop was a series of discussion charges to the community. These charges were designed to 1) engage the community in discussion, modification and further development of the Draft Assessment Framework for biosignature assessment, and in particular to determine if a generalized framework that spanned many biosignatures and detection methods was a feasible goal, and 2. Start the development of an ideal biosignature reporting protocol, including identifying incentives and disincentives to that protocol, and 3. Kick-start the writing process needed to engage the community in the workshop report.

These discussion group charges can be summarized as follows:

● *Try to break the Draft Assessment Framework using your favorite biosignature example. Step 1. Map your biosignature example to the Draft Assessment Framework (45mins)*

*Step 2. Assess the applicability of the framework and identify key missing aspects of the framework or other issues (45 mins)*

- *Design an Assessment Framework that meets the needs of multiple targets, instruments, biosignatures. Record changes needed to the protocol and any questions for discussion with the larger group (1.5 hrs)*
- *Discuss solutions to a chosen problem or issue with the Framework identified in the previous discussion session (1.5 hrs)*
- *Start fleshing out the report outline for the section you are responsible for and organize a time on Wednesday for a 2 hr writing session to work on the workshop report*
- *For the reporting protocol: Step 1: Discuss attributes of an ideal reporting protocol (scientific and communication) and identify existing structures and incentives that support or work against those attributes (45 mins) Step 2: What changes to the incentive structure would be needed to support the ideal reporting protocol?(45mins).*

Breakout group participation was either assigned based on participant measurement technique, selected by the participant, or assigned randomly depending on the needs of the breakout question. All material from the discussions in the US and satellite sessions was recorded in breakout-specific group notes in google docs. In the last 15 minutes of each breakout session, the participants would discuss and draft a single slide in a group slide presentation that was then used in the report out to plenary.

**The workshop report writing activities**
By the end of the workshop several participants had organized into groups and had already started writing parts of the workshop report outline. After the workshop, all discussion participants were invited to continue to help with the writing, and of the 130 discussion participants, roughly 50% (74 people) agreed to continue to help with the writing effort. These participants divided into groups addressing the six whitepaper sections (motivation, generalized framework, framework worked examples, framework applications to mission lifecycle, reporting protocol and future work). Writing meetings were held once every 1-2 weeks by most of the writing groups throughout August and September, and the writing teams and co-leads completed a draft on October 18th. This was released first to all the workshop participants (and all in the workshop Slack space, which included asynchronous attendees) for 10 days for comment, and then released to the general public on October 28th. The workshop participants left comments on the google doc which were addressed by the writing teams and co-leads. The public comment period is managed through a website interface where people can leave comments: https://www.nfold.org/soe-endorsements. 170 members of the community left comments and/or co-sign the workshop report. The comment period closed on November 18th.

Key dates for the reporting activity are as follows:

**Aug-September: Post-meeting writing sessions.**
**October 27: Finalize workshop whitepaper.**
**Oct 28 – November 18: Community comment period**
**November 10: Present to NAS CAPS**
**November 15: Present to NASA PAC**
**Early January: Finalize manuscript**
**January: Submit to NASA**
**April 2022: Submit for publication**
**Acronyms**

ALMA: Atacama Large Millimeter/submillimeter Array

CERN: Conseil Européen pour la Recherche Nucléaire , or European Council for Nuclear Research

CIA: Collision-Induced Absorption

CMOLD: Complex Molecules Detector

CNRS: Centre national de la recherche scientifique (French National Centre for Scientific Research)

COVID-19: Coronavirus Disease 2019

DAVINCI: Deep Atmosphere Venus Investigation of Noble gases, Chemistry, and Imaging

DL: Deep Learning

DLR: Deutsches Zentrum für Luft- und Raumfahrt (German Aerospace Center)

ESA: European Space Agency

ETH: Eidgenössische Technische Hochschule

ETI: Extraterrestrial Intelligence

ExoPAG: Exoplanet Exploration Program Analysis Group

FDA: U.S. Food and Drug Administration

GCMS or GC-MS: Gas Chromatograph Mass Spectrometer

GISS: Goddard Institute for Space Studies

GSFC: Goddard Space Flight Center

HabEx: Habitable Exoplanet Observatory

HITRAN: High Resolution Transmission

HZ: Habitable Zone

InSAR: Interferometric Synthetic Aperture Radar

IPCC: Intergovernmental Panel on Climate Change

ISO: International Organization for Standardization

JAXA: Japan Aerospace Exploration Agency

JPL: Jet Propulsion Laboratory

JWST: James Webb Space Telescope

LATMOS: Laboratoire atmosphères, milieux, observations spatiales

LCMS or LC-MS: Liquid Chromatography–Mass Spectrometry

LCROSS: Lunar CRater Observation and Sensing Satellite

LDKB – Life Detection Knowledge Base

LHC: Large Hadron Collider

LIFE: Large Interferometer For Exoplanets LPI: Lunar and Planetary Institute

LLC: Limited Liability Company

LR: Labeled Release

LSW: Landessternwarte Königstuhl

LUVOIR: Large Ultraviolet Optical Infrared Surveyor

MEPAG: Mars Exploration Program Analysis Group

MISS – Microbially-Induced Sedimentary Structures

ML: Machine Learning

MTBSTFA: N-Methyl-N- (Tert-Butyldimethylsilyl)trifluoroacetamide

NAO: National Astronomical Observatory

NASA: National Aeronautics and Space Administration

NASEM: National Academies of Sciences, Engineering, and Medicine

NExSS: Nexus for Exoplanet System Science

NfoLD: Network for Life Detection

NIH: National Institutes of Health

NMR: Nuclear Magnetic Resonance

NOW: Network for Ocean Worlds

NSF: National Science Foundation

OPAG: Outer Planets Assessment Group

ORAU: Oak Ridge Associated Universities

SAM: Sample Analysis at Mars

SETI: Search for Extraterrestrial Intelligence

SNR – Signal-To-Noise Ratio

STM: Science Traceability Matrix

TDEM: Technology Development for Exoplanet Missions

TRAPPIST: Transiting Planets and Planetesimals Small Telescope

UC: University of California

UCLA: University of California, Los Angeles

UNESCO: United Nations Educational, Scientific and Cultural Organization

UPR: Universidad de Puerto Rico

US: United States

UV-VIS-NIR: Ultraviolet-Visible-Near-Infrared

VERITAS: Venus Emissivity, Radio science, InSAR, Topography, And Spectroscopy

VRE: Vegetation Red Edge

## 1. Are we alone?

Astrobiology is the study of the origin, evolution, distribution and future of life in the universe, and one of its ultimate goals is the detection of life beyond Earth. However, the complexity of life itself, the scientific challenges of the required measurements, and the difficulty in conveying this complexity to others, means that it is now critical for the astrobiology community to develop standards of evidence and reporting for the detection of potential signs of life. A claim of life detection is a high-stakes scientific achievement that will garner enormous public interest, yet there is a high potential for any initial life detection claim to be ambiguous. Current and upcoming research efforts and missions aimed at detecting past and extant life could have their goals supported by a consensus framework to plan for, assess and discuss life detection claims (cf. Green et al., 2021) that minimizes the chance of false positives and ensures clear communication with the scientific community and the public.

### 1.1 Ongoing and future missions will enable the search for life beyond Earth

In the past two decades, extraterrestrial life detection has become a more explicit and central goal for both planetary science and astronomy research objectives and missions. To have maximum applicability to current and future missions that search for life, any biosignature assessment framework needs to address discoveries from both missions that travel to the celestial bodies within our Solar System---that can collect orbital, fly-by or *in-situ* data, and missions that search for signatures of life on extrasolar planets via remote observation. This is not a trivial task, as both the signatures of life and the sources of false detections are strongly dependent on the source of the data itself, and the context of the environment in which it is obtained. In the coming decades, the developed framework will need to be applicable to searches for life on the following astrobiological targets, to which missions are either in progress or being planned.

### 1.1.1 Missions on Local Terrestrial Worlds:

Our terrestrial neighbors are of high astrobiological interest due to their relative accessibility, and the potential for comparison with Earth-based analog studies. The latter is key to how we learn to study life, and any framework for life detection must also apply to Earth-based studies, as well as extraterrestrial data. Mars has been a primary candidate for life beyond Earth, since the planet is likely to have possessed liquid water and broadly habitable surface conditions in its early history, and may even provide potential subsurface refugia for life today, including subsurface liquid water (e.g., Des Marais et al., 2008). The NASA Mars 2020 Perseverance rover is investigating evidence for past life in Mars's Jezero crater, and has recently collected samples for future transport back to Earth for detailed study including the search for biosignatures (Williford et al., 2018). The ExoMars program will also search for signs of past and present life on Mars, and includes a currently operational orbiter investigating Martian atmospheric trace gases and their sources. This spacecraft will interface with the Rosalind Franklin rover (planned to launch in 2022) that will also search for water, variations in the geochemical environment, as well as organic molecules that could be indicative of past or present life on Mars. Venus is another terrestrial body of interest that may have been habitable in the past. Two NASA missions, VERITAS and DAVINCI, and ESA's EnVision mission are planned for launch in the next decade, and will provide new insight into the past and present environment of Venus, including the detailed composition of the current atmosphere and insight into potential false-positive biosignatures (e.g., Meadows, 2017).

### 1.1.2 Missions on Local Ocean Worlds:

The icy-moon ocean-worlds that orbit Jupiter and Saturn, such as Europa, Enceladus and Titan, have emerged as important targets in the search for life within the Solar System due to the availability of liquid water beneath their icy surface, and sources of energy (Hendrix et al., 2019). NASA's Cassini Saturn orbiter also detected the presence of organics on Enceladus and Titan (Postberg et al., 2018). A variety of missions have been selected and proposed to explicitly study the astrobiological potential of Europa (Europa Clipper: Pappalardo et al., 2017; Hand et al., 2017), Titan (Dragonfly: Turtle et al., 2019) and Enceladus (MacKenzie et al., 2021; Cable et al., 2016). Also, ESA has recently selected as a top priority the investigation of the habitability potential of the moons of the giant planets for future large-class missions in the timeframe 2035-2050.

### 1.1.3 Exoplanetary Missions:

In the coming decades, astrobiological studies outside our Solar System will gather data from remote observations using either ground-based telescopes and instruments or those in space. Characterizing the atmosphere and surface of potentially habitable planets and searching for signs of life via transmitted, reflected and emitted radiation is enabled in the next decade by current and near-term ground and space-based telescopes, including the James Webb Space Telescope (e.g., Krissansen-Totton et al., 2018), and the next generation of Extremely Large Telescopes (ELTs; e.g., Lopez-Morales et al., 2019). Future proposed missions that would enable more comprehensive searches for habitability and life on exoplanets include the NASA HabEx (Gaudi et al., 2020), LUVOIR (The LUVOIR Team, 2019) and Origins (Meixner et al., 2019) concepts, and ESA's LIFE mission concept (Quanz et al., 2021).

### 1.2 Past claims of life detection inform the processes required for biosignature assessment

While ongoing and future missions are poised to expand the search for life elsewhere, existing and past claims of possible biosignature detection within the Solar System have demonstrated the complex nature of the life detection process, and the community-wide efforts needed to assess these claims. The ensuing debates, within the scientific community and the public at large, also illustrate the necessity for an assessment framework to systematically increase the robustness of the life detection claim being made, and a reporting protocol that could communicate the existing evidence and the assessment process more accurately, both within the field and to a wider audience.

One of the most famous claims for the possible detection of extant life on Mars (Levin and Straat, 2016) is based on the results of Labeled Release (LR) experiment carried on the NASA Viking Landers (Levin and Stratt, 1976a). The LR experiment tested for the presence of life by adding carbon-14 labeled ($^{14}$C) organics to Mars regolith in a sample cell while monitoring for production of $^{14}$C-gas that might be indicative of microbial metabolism, and in some respects the results resemble what had been expected for a biological response (Klein et al., 1976; Levin and Stratt, 1976b; 1977; 1979).  However, since Viking it's been recognized that there are numerous possible non-biological explanations for the LR results (e.g., Ponnamperuma et al. 1977; Oro, 1979; Plumb et al. 1989; Quinn & Zent, 1999). The results of the Viking Gas Exchange Experiment, which demonstrated that the regolith was chemically reactive (Klein et al. 1976, Oyama et al., 1977; Oyama and Berdahl, 1977), and the search for organics using the Viking Gas Chromatograph-Mass Spectrometer (Biemann et al., 1976; Biemann et al., 1977; Biemann & Lavoie, 1979) further support a non-biological interpretation of the LR results. Additionally, detections of oxychlorine compounds on Mars (perchlorate and chlorate) during the Phoenix and Mars Science Laboratory missions (Hecht et al., 2009; Glavin et al., 2013) indicate the occurrence of processes that can explain both the Viking LR and GCMS results (Navarro-Gonzàlez et al., 2010; Quinn et al., 2013; Guzman et al., 2018). Although systematic reviews of the Viking results have been performed to evaluate whether life was discovered (e.g., Klein, 1978; 1999) they have been performed outside of a standing community accepted framework for evaluating life-detection claims. The decades-long debate over the LR results clearly demonstrates the need for such a framework.

Claims for traces of past life in the martian meteorite ALH84001 is another example of a life detection claim that was later found to be more likely due to abiotic processes and characteristics. The McKay et al. (1996) team based their claims for the possible presence of life in the meteorite on multiple lines of evidence ranging from low temperature fluid infiltration of fractures in the cumulate volcanic rock, globular to chain-like physical structures in carbonate deposits identified as "nanobacteria", nanocrystals of magnetite identified as magnetosomes, and the presence of organic matter in the form of polycyclic aromatic hydrocarbons. However, at that time general understanding of the microbial world and how it is preserved in the rock record, as well as the potential for abiotic processes to create signatures that can imitate biogenic ones, was significantly limited. Today, this claim is discounted, although the very fact that a workshop on standards of evidence for life detection was held hails back to this landmark paper.

More recently, claims of phosphine detection in the Venus atmosphere (Greaves et al., 2020), sparked a response that illustrated the multi-step process and necessity for community involvement in comprehensively assessing potential biosignatures. Although apparently detected in several observations with two telescopes, multiple sets of researchers reanalyzed the discovery data and questioned the significance of the detection (Snellen et al., 2020; Akins et al., 2021; Villanueva et al., 2021; Thompson, 2021). Others searched for additional features from phosphine at other wavelengths or with different instruments, in an attempt to corroborate the initial detection (Encrenaz et al., 2020; Trompet et al. 2020; Mogul et al., 2021). Modeling of the observed data suggested that the signal was more plausibly due to sulfur dioxide, a common gas in Venus' atmosphere which also absorbs near the same frequency position, rather than $PH_3$ (Lincowski et al., 2021; Akins et al., 2021; Villanueva et al., 2021), although this too is now contested (Greaves et al., 2021). While the discoverers performed an exhaustive survey of potential abiotic production mechanisms but could not identify a likely source (Greaves et al., 2020; Bains et al., 2021), others worked to find plausible abiotic formation mechanisms for $PH_3$ in the Venus atmosphere (Truong & Lunine 2021; Omran et al., 2021). This rapidly evolving scientific story illustrates the progressive way in which scientists combine a series of results to more accurately assess the detection of possible biosignatures.

Claims of signs of life in the early rock record similarly illustrate the need for consistent criteria in determining biological provenance, as well as the importance of contamination evaluation and the enhanced rigor enabled when multiple teams engage in assessment of claims of biosignature detection. In one example, Schopf (1993) described "microfossil-like objects", which he interpreted as eleven different microbial taxa in a ~3.5 Ga sedimentary chert rock. These structures were reinterpreted by Brasier et al. (2002) as abiotic organic matter in a hydrothermal setting. The debate has continued for over 30 years, with Rouillard et al. (2021) noting that the criteria used for biological origin are inconsistent within the community, causing the fossil-like structures to sometimes be presented as evidence for biological activity, and other times as abiotic. This example underscores the need for community-wide adoption of a standard set of criteria for assessing potential preserved biosignatures. In another example, Brocks et al. (1999) identified hopane and sterane organic compounds in Archean sedimentary rocks, which can be indicative of bacterial and eukaryotic cells, respectively, having used the best protocols at the time to rule out contamination. This research led to the interpretation that eukaryotes flourished 300 million years prior to the oxidation of the atmosphere, a view that was supported for almost a decade until Rasmussen et al. (2008) showed that while organic matter was indigenous to the rock, the hopane and sterane biosignatures were likely, in fact, the result of contamination. This was confirmed by French et al. (2015) who conducted a drilling campaign that was specifically designed to further minimize contamination, followed by analysis in multiple laboratories utilizing clean procedures. This example also demonstrates the importance of a rigorous and robust multi-laboratory approach in increasing the confidence in biosignature detection.

### 1.3 The path to life detection

Given the examples above, it is clear that the detection of extraterrestrial life is unlikely to be instantaneous or unambiguous, and yet it is a high-stakes scientific achievement that will garner enormous public interest. Even when examining evidence of early life on Earth under the most ideal analytical conditions, detecting and interpreting biosignatures is extremely challenging, requiring multiple measurement techniques and contextual information, sometimes developed over decades. For the arguably more ambitious goal of detecting extraterrestrial life, we will likely embark on even longer journeys involving a series of measurements and missions, each progressively increasing our confidence in the interpretation of our findings, and helping to provide key aspects of biosignature detection, confirmation and assessment. It is crucial for our field to clearly convey the complex, multi-measurement assessment strategy necessary for biosignature detection. Such a framework will allow us to articulate the inherent value of a measurement or mission in increasing our confidence in life detection, and how it advances us on the path to life detection, even when that mission is not returning definitive evidence of life itself.

With claims of possible life detection becoming more frequent, it is critical to discuss as a community how to agree upon a universal scientific framework to best vet and support discoveries, and to report and convey information on a topic that is inherently complex and interdisciplinary. Green et al. (2021) articulated one such example framework for biosignature assessment and encouraged the astrobiology community to refine and develop their own. The Biosignature Standards of Evidence (SoE) Workshop took up that charge, and additionally discussed how to best communicate claims of life detection to our scientific colleagues and to the public.

In the remainder of this document, we present a community-developed, universal framework that can be used to improve assessment protocols for biosignature detection claims, contextualize a measurement or mission result, and more accurately communicate confidence in results related to life detection. Construction of such a framework is not trivial, as different fields will have inherently different ways to assess their confidence in biosignature detection and assessment, and there is an inherent tension between a rubric that is simple to communicate and use across a wide range of life detection scenarios, but nuanced enough to handle the complex physical and chemical phenomena measured using a wide range of techniques. In particular, investigations in our own Solar System generally allow for the collection of samples or analysis *in situ*, yet the search for life on exoplanets depends on telescopic measurements from great distances. Reconciling the standards of life detection between these two communities is complex, but crucial. With this in mind, the framework described through this paper was developed via a massive interdisciplinary collaboration from the different fields within the astrobiology community, and the hope is this will be continuously improved upon and developed over time as we progress in our search for life.

In this workshop report, Section 2 outlines the structure of the framework itself. Section 3 provides historical and hypothetical examples of claims of possible biological activity to illustrate how the framework might work in practice for a range of targets and measurement techniques. Section 4 explores deeper applications of the framework to multiple phases of mission development and lifetime. Section 5 describes a protocol for reporting biosignature assessment results. Section 6 outlines possible pathways of research that would help to fortify the aims of this framework.

## 2. Towards A Generalized Framework for Biosignature Assessment

In this section we describe the first steps in developing a generalized framework for biosignature assessment. As motivated by Section 1, to scientifically assess and accurately communicate likelihood in claims of life detection, there is a need for a community framework that identifies key questions and provides a generalized description of the scientific processes and evidence needed for this assessment. We use the term *'generalized'* here to denote that the ideal framework is sufficiently broad and process-based that it can be useful when applied to life detection claims made using a variety of measurement techniques, including both *in situ* and remote-sensing, for a variety of targets within and beyond our Solar System, and for extinct and extant life. This framework could be used either once a claim of biosignature detection is made, or proactively, to justify the efficacy of measurements and theoretical work in enhancing future life detection efforts. In this section, we outline the assessment framework structure that was developed and refined at the Biosignature Standards of Evidence workshop, and describe in more detail the generalized scientific questions to be answered in each step.

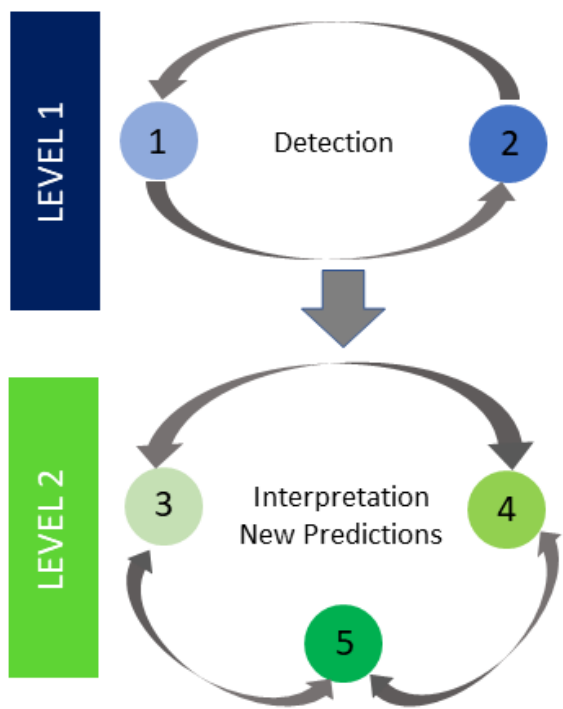

*Figure 2.1: Schematic of the Community Biosignature Assessment Framework. Level 1 encompasses Questions 1 and 2 to be addressed to increase certainty in potential biosignature detection and identification, and Level 2 includes Questions 3-5 centered on assessment and discrimination between abiotic and biological processes as the source of the potential biosignature, and may generate new predictions needed to enhance confidence in life detection.*

## 2.1 Overview of the generalized framework

The assessment framework comprises five Questions clustered into two broad components (Figure 2.1): Level 1, which assesses our confidence in the detection and identification of a potential biosignature, and Level 2, which assesses whether or not the potential biosignature is more or less likely to be due to life, especially in the context of the environment in which it is found. More specifically, Level 1 of the framework encourages those who discover a potential biosignature to address Questions 1 and 2, which are: "Have you detected an authentic signal?", and "Have you adequately identified the signal?". Level 2 moves from detection and identification, to assessment of the likelihood that the signal is due to life and includes Questions 3, 4, and 5: "Are there abiotic sources for your detection?"; "Is it likely that life would produce this expression in this environment?"; and "Are there independent lines of evidence to support a biological (or non-biological) explanation?".

## 2.2 General considerations when applying the framework

While the framework questions are presented in order, there was a strong sense at the workshop that the application of the steps do not need to be in a particular order, and in some cases may be difficult to implement linearly. Within the levels, measurements and processes may be cyclical or iterative, as a given measurement or modeling result might simultaneously address several questions, and further measurements or other research to address subsequent questions may in fact reduce, rather than enhance, the degree of certainty in a given interpretation. However, to minimize impact on the broader community from a biosignature detection claim, it is considered desirable to have high confidence in Level 1 (the biosignature candidate detection and identification) prior to moving on to Level 2, where significant community efforts will be needed to assess whether a biological source is a likely explanation for the potential biosignature.

The assessment framework is also designed to be generalized, and therefore applicable to a broad range of biosignature types and measurement techniques. While the specific application of the framework and the assessment of certainty in a given measurement or interpretation will vary as a function of measurement technique, target and biosignature sought (see Section 3 for more detailed field-specific examples), the questions that we ask ourselves as astrobiologists---and the processes we must perform to enhance scientific robustness---are more universal. For example, confidence in a biosignature detection is deeply dependent on an understanding of instrumental systematics and biases, a process common to all, but for which the specifics vary depending on instrument and field. Similarly, confirmation via multiple measurements or additional instruments is considered more robust than a single measurement alone, but which instruments could be brought to bear is very field and biosignature specific.

Moreover, in grasping and applying this framework, challenges can arise in the different terminology used by different fields for similar concepts, or for different processes that have similar impacts on our ability to detect or assess biosignatures. For example, both coelution of compounds in a chromatographic sample, and blending of lines or bands from two different molecular absorbers in the spectrum of a planet, make the unique identification of a potential biosignature more difficult. Even though the techniques and targeted biosignatures are extremely different, the process of disambiguation of these measurements is a universal step that would be required to increase the robustness of biosignature detection and assessment. Consequently, it is expected that although the biosignature assessment framework contains generalized questions and processes, its application will be different and adaptable to the needs of a given technique and its targeted biosignature.

Other discussions around applying a more generalized framework also centered on the recognition that addressing certain questions may be more challenging, and require a different threshold for confidence depending on the investigation type, whether that be *in situ* or remote, ancient vs. extant life, or even evidence of life that does not assume the same heritage and biochemistry as we observe on Earth. To help maintain a generalized framework, it was considered valuable that these types of measurements should be assessed on the same confidence scale, even if very different certainties were considered achievable due to inherent limitations in the measurement technique or data sought. Discussions were initiated on the difficulty of assigning numerical confidence to a given question, especially across fields, and whether a more simplistic three-level scale might be more appropriate. Combining different confidence levels for different questions into a single confidence value was also discussed as a possibility, as was at least marking when a given study had cleared Level 1 to the community's satisfaction.

## 2.3 The Generalized Framework for Biosignature Assessment

Below, we describe the framework questions in detail, providing a plain language summary of the intent of the question, the motivation to address the question, and known techniques and scientific strategies for how we might best address the question.

### Question 1: Have you detected an authentic signal?

*Have you authenticated your signal, and is it statistically significant? Have you ruled out artifacts from the measurement, pre-processing and/or analysis process that might mimic a real signal?*

A claim of biosignature detection begins with an observation or a measurement of a particular phenomenon, a signal, which may be anything from structures observed in *in situ* collection of sample data, to a remotely sensed molecular absorption. The first assessment question is then "is this a real signal?". Answering this question first is critical, because if there is a low probability that the signal is real, then biosignature assessment does not need to proceed to subsequent steps. Determination of whether the observed/measured signal is real will rely on the field-accepted standards of measurement for the relevant instrument, and its associated measurement and processing techniques. Reasons for which the signal may not be real include the injection and/or misinterpretation of noise or contamination due to an instrumental, measurement technique or processing artifact. For example, in planetary spectroscopy, reflections in the instrument or other noise sources may produce an apparent signal, where in reality none exists. In the laboratory, extracting organics with hot liquid solvents may generate molecules that did not exist in the original sample. The key guiding question then becomes, what unknowns, noise, measurement technique, instrumental or analysis limitations must be addressed to consider the detection legitimate?

There are several procedures and techniques that can be used to increase confidence that a given biosignature detection represents a real signal. The signal characteristics should be of adequate quality, including signal-to-noise ratio (SNR) and measurement frequency. *A priori* knowledge of instrument performance and limitations through validation is an important factor, as it helps to identify previously known spurious signals from the instrument, and also helps to determine if the reported signal is within the instrument's expected sensitivity and performance. Calibration blanks or other reference measurements used at the time of, or in conjunction with the measurement, can be used to rule out spurious instrumental or measurement system signals. These validations and calibrations, as well as other key aspects of the way samples are taken, or the way the phenomenon is measured, should be used to estimate the uncertainties and the corresponding detection significance using standard field-specific techniques. Confidence in the signal can also be increased via multiple detections on the same sample, potentially with multiple instruments that provide convergent data, and if possible on different samples. Similarly, reanalysis of the sample or detection data by multiple independent teams can be used to strengthen confidence in the detection. Data processing and statistical analysis techniques, such as resampling via bootstrapping or using other perturbations can also be brought to bear to test the stability of the signal, and estimate true confidence intervals for the likely range of values associated with the detection. The repetition of the detection with temporally-spaced samples may also increase confidence that it was not due to transient noise, as long as the phenomenon is expected to persist over time, and that any spurious instrumental signals or contamination is not.

## Question 2: Have you adequately identified the signal?

*Have you adequately ruled out other potential sources for this signal? For example, have you ruled out contamination in the environment, or other real phenomena that could produce a similar signal?*

Having established a high degree of confidence that the potential biosignature signal is an observed phenomenon in the sample or planetary target, and not due to instrumental, measurement technique or processing artifacts, Question 2 then addresses whether the observed signal has been adequately identified, and attributed to the correct phenomenon. Question 2 therefore focuses more on issues of misidentification of the biosignature due to contamination via phenomena that do exist elsewhere in the environment or the sample, but are not the target biosignature sought. Like Question 1, addressing this Question increases the likelihood that a potential biosignature has indeed been detected, prior to the more comprehensive and interdisciplinary effort outlined in Questions 3-5 that are needed to interpret that biosignature detection. In some fields, addressing Questions 1 and 2 may need to be done, or naturally occur, simultaneously, rather than as separate steps. Specific examples of possible phenomena that may confuse biosignature identification include misidentification or coelution of spectral or chromatograph lines from multiple molecules absorbing at the same position, or contamination of rock samples with non-indigenous organics. The key guiding question becomes: can you be reasonably sure that the observable signal is representative of the molecule, process, or other line of evidence to which it is attributed, and for the sample or target under consideration, and is not due to another phenomenon?

There are multiple techniques and procedures that could be used to enhance confidence in the correct identification of the signal. Understanding the limits of detection or resolution for the observed signal can be a first step toward determining whether a misidentification is likely. Other steps to confirm identification can include comparing the measurement with known measurement-specific standards or libraries, which can also be used to reveal potential confusion with other phenomena. Steps can then be taken to account for contamination of the sample or other interactions with the environment. Multiple samples can be taken and compared to look for consistency. Similarly, comparisons with procedural or instrument "blanks" can reveal contamination. Sample history, including determination of syngeneity (that the age of the rock and any putative biological material within is equivalent; Oehler & Cady, 2014) and spatial relationships to structures in the sample, can help discriminate contamination by more recent biological material. Careful site selection for sampling, and recording of sampling context can also help with contamination control and identification. Obtaining additional, complementary measurements and contextual clues to break any potential degeneracies can include confirming the identification with multiple instruments and/or features.  For example, searching for additional absorption features from a molecule in spectra taken with different instruments at different wavelengths, or using multiple techniques on a single rock sample to provide corroborating isotopic, compositional and spatial information to identify embedded organics. Finally, contextual clues from the environment may provide further support for the identification by showing that the detection is consistent with other physical, chemical, and/or geological parameters detected or characterized at the site.  Although both Question 2 and Level 2 Questions (Q3-Q5, below) rely on environmental context, this information in used in Question 2 is to identify the signal whereas, the in Question 3 it is used to identify the process generating the signal. In some cases measurements made to address Question 3 may be necessary to strengthen identifications made for Question 2.

## Question 3: Are there abiotic sources for your detection?

*Is it likely that there is a current or past environmental process, other than life, that could be producing this signal? Have you ruled out potential false positives for the biosignature?*

Once the detection of a signal has been confirmed and possible contamination or misidentification has been addressed as rigorously as conditions allow (Questions 1 and 2), the assessment framework moves towards the steps needed to determine whether or not the signal is due to a biological process (Questions 3-5). The first of these steps (Question 3) is to determine if there are any plausible abiotic sources of the signal, and if so, to attempt to rule them out. This is a critical first step in assessing the biogenicity of a signal as it conservatively assumes that the signal is not due to life. If a convincing abiotic explanation is found, it can reduce the probability that a true biosignature has been observed. However, as each alternative abiotic explanation is proposed and then potentially ruled out, it can also strengthen the case---although not prove---that the biosignature may be due to life. While a biosignature by definition reveals life's impact on its environment, our interpretation of the biogenicity (or abiogenicity) of a detection will in turn strongly depend on the (paleo)environmental context. A recognized biosignature in one environment may not directly translate to another, and abiotic planetary processes in a different environment may produce "false positives" for the biosignature. Examples of proposed false positives for putative biosignatures include: modeling of the production of atmospheric $O_2$ via ocean loss and photochemistry, rather than photosynthesis or other biological processes (cf. Meadows, 2018, and references therein); laboratory generation of prebiotic compounds and synthetic organic chemistry (Barge et al., 2021); observations of abiotically sourced molecules detected at hydrothermal vents (Ménez et al., 2018); measurements of amino acids and other structurally intricate organic compounds in meteorites (Burton et al., 2012; Glavin et al., 2020); and measurements of aromatic organics generated during testing of Martian samples by the Mars Science Laboratory (Curiosity Rover) (Szopa et al., 2020).

Key challenges in identifying abiotic processes for *in situ* and early Earth studies include: limitations to our understanding of how geochemical systems evolve over time; the preservation potential of abiotic signatures; the historical focus on simulating prebiotic chemistry in early Earth environments rather than conditions relevant to other worlds; technological constraints in characterizing organic content on other planetary bodies; and, of course, the potential sheer enormity of environments and processes that may produce abiotic signals. For exoplanet studies, the diversity of potential planetary environments and processes is likely even larger, and must be explored solely via modeling until observational data on exoplanet environments becomes available. However, our understanding of terrestrial exoplanet environments can be informed by geological and atmospheric planetary processes from planets in our Solar System (Kane et al., 2021), as well as stellar/planetary/planetary system interactions.

However, despite these challenges, there are some potential best practices. Published methods describing abiotic sources for the potential biosignature, or chemically similar compounds, should be sought. This includes whether or not a molecule is detected to high confidence in astromaterials or the interstellar medium, or has been generated in laboratory analog or prebiotic chemistry studies. If an abiotic synthesis mechanism is known, then other molecules or signals that might be part of that mechanism should be identified and sought to help confirm or exclude abiotic generation. The presence of mineral or metal catalysts should be noted for their potential to drive additional chemistry but also with the understanding the potential reaction possibilities for these metals given that there is coevolution between geochemistry and biochemistry (i.e. metallo-enzymes). The stability of the molecule under the conditions in which it is found should also be considered, including whether its abundance changes over time. An active flux may not always be an indicator of life, as it could be an abiotic release mechanism, although potentially sourced from a preserved residual component of extinct life. Alternatively, when looking for signs of life in remote-sensing data, an active, potentially changing flux of a molecule (e.g., seasonally) may be biological and so requires further investigation. The stability of the molecular isomers, enantiomers, and homologs should also be considered and can aid in the evaluation of the detection being sourced via abiotic mechanisms. In particular, whether abiotic sources consistently lead to the production of homologs and isomers of the proposed biosignature should be assessed. If so, their absence (if the technique was able to detect them) would build a stronger case against an abiotic origin; if homologs or isomers could not be detected or resolved with the given detection method, this should be noted. Long-term resilience or "preservation potential" of a biosignature is a key parameter (along with origination/plausibility and concentration) for a reliable biosignature interpretation (cf. Westall et al., 2018).

Because abiotic signatures may have been generated earlier in the environmental lifetime, and been preserved, mechanisms that may have operated under conditions different to the present-day environment must also be considered. For morphological signals, considerations of whether or not the signal is widespread, and the degree of degradation, will be important considerations for assessing whether an abiotic source for the structure is more likely.

Finally, for consideration of abiotic processes in assessing the biosignature, we must evaluate the state of community knowledge pertaining to those processes or classes of processes. If the only proposed abiotic model or mechanisms are poorly characterized, or if the scope of possible abiotic explanations is known to be poorly explored, it suggests we cannot adequately reject abiotic mechanisms. Any detection claim should include the identification of areas of further study that would be required to identify or preclude possible abiotic sources. This information should be considered in combination with the analysis in Question 4, on the likelihood of the life hypothesis.

### Question 4: Is it likely that life would produce this expression in this environment?
*Given what we know about the likely environment that an organism is operating in, or would have operated in, does it make physical and chemical sense that life would produce this potential biosignature?*

Critically, life detection claims should not only rest on ruling out abiotic sources, but also on the viability of biological explanations. A claim of extraterrestrial life detection is extraordinary, and this hypothesis should be rigorously tested in its own right, and not simply considered the most likely explanation once all known potential abiotic mechanisms have been sought and excluded.

For a signal to be a biosignature, it must be consistent with the products or activity of life whose nature can be supported by the (paleo)environment in question. Addressing Question 4---and making the case for the likelihood that life could have produced the signal in the environment under consideration---therefore complements efforts in addressing potential abiotic sources in Question 3, and both activities work towards attempting to support or rule out a biological explanation for a biosignature. Key considerations relevant to answering Question 4 include determining whether or not the environment in which the candidate biosignature has been found is currently or was previously habitable, that is, whether it contains the resources and physicochemical conditions required to support life, or whether it could have potentially supported life in the past. This interpretation should also consider that the environment in which the biosignature was synthesized may not be the environment in which it was measured, due to environmental evolution and/or transport processes. A justification also needs to be developed as to why life would make this signal in the observed environment. This justification may involve comparison with the known behavior of Earth organisms in similar environments, or may require the development of a new model of life that is consistent with what is known of the non-Earth-like environment. Important considerations in developing such a model include whether or not the production of the biosignature is physically or chemically plausible in the environment. For a biosignature that may have been generated by ancient life, the likelihood of signal preservation also needs to be considered.

A habitability assessment is a key component of determining whether it is likely that life can produce (or could have produced) a biosignature in the environment under study. In many cases, an initial habitability assessment will already have been performed for site or target selection before the biosignature measurement was made. These assessments are necessarily bounded by what we know about the limits to life on Earth. Where these limits have been exceeded in a specific environmental parameter, the biosignature could, depending on context, be rejected. Alternatively, further analysis could be undertaken to discover if that particular parameter measurement is reliable, or could plausibly be accommodated by biochemistry. The detection of a potential biosignature may spur further environmental characterization on multiple spatial and temporal scales, in order to improve our ability to identify false positive mechanisms (see Question 3) and to assess the likelihood that life could exist in that environment.

Attempts to justify whether life might produce the detected signal in the measured environment can also consider if an assumed model of life's impact on its environment is physically and chemically consistent with the phenomena observed, and consistent with what is known of the environment. Current bold goals include progress towards models that can determine if it makes environmentally contextual sense that life produced the observed phenomenon can be made without a definitive definition of life. If life on Earth is known to produce similar signals in analog environments, then a direct comparison can be made. However, if the extraterrestrial environment has no Earth analog, a model of how life—including the possibility of life as we do not know it—might function in that non-Earth environment, and plausibly produce the detected potential biosignature, may need to be proposed. For example, if the potential biosignature is proposed to be a metabolite, then assessment of whether life has a thermodynamic impetus or benefit in creating such a product in that environment would need to be developed.

There are a number of ways that the plausibility of life producing an environmental phenomenon could be assessed. For example, plausibility arguments can potentially be made by looking at the abundance or inferred flux of the potential biosignature and checking whether it is consistent with estimates of the free energy or nutrient availability in that environment, and whether it is out of equilibrium with its environment. For biosignatures that are characteristic physical patterns, molecular distributions, or complex signals or molecules, probabilistic arguments can be made for biological processes being more likely than abiotic ones to produce these features. For biosignatures of fossilized or remnant life, a further plausibility consideration is whether it is likely for the observed signal to have been preserved over geologic time in this environment, or be more likely due to forward contamination. Beyond determining if the potential biosignature "makes sense" given the environment in which it is observed, broader considerations of the plausibility that life would produce this signature include considerations of the probability that life emerged on the body in question. However, this requires the incredibly challenging acquisition of constraints on the environment in which life likely evolved, which may be very different to the environment under study. This ancient context is tied to the planet's evolutionary history, which may be somewhat knowable (e.g., for Mars) or extremely ill-defined (e.g., for exoplanets).

## Question 5: Are there independent lines of evidence to support a biological (or non-biological) explanation?

*Are there other measurements that provide additional evidence, or allow you to predict and execute follow-on experiments, that will help discriminate between the life or non-life hypotheses?*

Although both the hypotheses of non-life and life will have already been tested in Questions 3 and 4, it is likely that the question of the presence or absence of life will still be unresolved and will require ongoing iteration. Addressing Question 5 is intended to provide additional rigor to the investigation by encouraging the further acquisition of multiple independent lines of evidence, and by the prediction and testing of hypotheses in support of either the biological or non-biological explanation for the observed potential biosignature. Developing theories for how life may be generating the signal, including characterizing the environment in which it is found, requires constantly improving currently incomplete models of geophysics, geochemistry, biochemistry, taphonomy, and many other disciplines, and so single lines of evidence are anticipated to be inadequate for assessing purported biosignature detection reports. The gold standard of any life detection study is the acquisition of multiple lines of evidence. If these have not been obtained, then predictions of other features or phenomena that could confirm the theoretical framework used to asses Questions 3 and 4 (see Green et al., 2021), and follow-up experiments to search for these phenomena, are needed.

Further assessment of biosignature interpretation could include historical research in the literature, reanalysis of existing data by multiple teams, and the development of new follow-on experiments. A thorough check should be made to determine if previously published research supports or contradicts the hypothesis to be tested, and multiple existing studies could be analyzed to better understand reproducibility (Botvinik-Nezer et al., 2020).

Best practices would encourage authors to make their methods and data readily available so that other groups can replicate results. When possible, independent labs and research teams should confirm potential biosignature observations. It is not necessary that results be replicated by multiple teams before publication of the discovery, but if results are not replicated, then the strength of conclusions should be moderated. It is important to note that multiple lines of evidence from different labs and/or research groups do not *guarantee* that the observation is a true, valid biosignature. However, given the importance of life detection claims, independent replication of observations and data analysis is crucial, since higher credibility will be afforded to multiple observations of the same biosignature, and/or a collection of potential biosignatures.

## 2.4 Future work needed to develop the Assessment Framework

The Biosignatures Standards of Evidence Workshop allowed us to make significant inroads in developing the above community consensus framework. However, further work, including potentially another workshop, will be needed to determine additional and field-specific details on how we can best assign confidence levels based on the data available, and whether a numerical, color, or other scale is the most useful in communicating degrees of certainty about the veracity of a given biosignature claim. Another potential avenue for further community discussion is the possible usefulness of combining responses to the questions into a single numerical or color ranking that conveys certainty in the current assessment of the biosignature claim. Such a task may be quite difficult to implement, however, and may need to depend on an external body for biosignature assessment.

Future work may also include a discussion on how best to use the framework in conjunction with the NASA Life Detection Knowledge Base (LDKB), which provides a community-driven assessment of potential biosignatures to be sought, and the potential for false positives and false negatives associated with specific measurements. The LDKB can serve as a central repository for community knowledge on how well studied a particular biosignature is, including its false positives and negatives, to aid in the selection of target biosignatures. The LDKB could also be used in conjunction with the biosignature Assessment Framework to inform the utility of specific life detection measurements in confirming the presence of life or revealing potential abiotic sources and potentially help with mission concept development (cf. Section 4).

Finally, the Framework can be used to inspire the laboratory, field and modeling communities to work towards comprehensively exploring plausible abiotic sources of a particular signal, determining the relevant follow-up signals necessary to distinguish abiotic cases, and ensuring that missions are designed with a biosignature assessment plan in mind, including considerations of how to rule out false positive scenarios. This also allows appropriate resources to be allocated to any particular biosignature within a mission (e.g., observing time to achieve desired signal-to-noise ratio in a given spectral band, or measurement of more isotopic ratios to increase precision), and identifies which mission aspects might benefit from additional resource allocation.

## 3. Worked Examples of the Assessment Framework

The five steps of the Assessment Framework were discussed at length over two days during the Standards of Evidence workshop, both within groups that focused on particular biosignature classes, disciplinary specialties, analytical targets, or instrumentation, and within groups that were purposefully randomized. All of these groups were asked to address two particular questions: Can the five steps of the Assessment Framework be mapped to the needs of your specialty? and Are there key missing aspects that would affect applicability? To summarize these activities, we have prepared four worked examples to demonstrate how the Assessment Framework can be used. We include an example from paleobiology, a discipline that has long grappled with distinguishing signs of life in the geologic record, an example from the analysis of organic matter on Mars to demonstrate an in situ application, a spectroscopic example that demonstrates utility for remote sensing applications, and an example that uses an "agnostic" lens on a possible biosignature without relying on known life on Earth for comparison.

## 3.1 Early Earth Worked Example: Detection and Characterization of Stromatolites

Stromatolites are laminated sedimentary structures that form at the sediment water interface in freshwater, marine, and evaporitic environments under the influence of growth or metabolic activity. This definition aligns with Grey and Awramik (2020), who consider stromatolites to be a subset of microbialite (for definitions of other types of microbialite see also: Riding, 2011; Bosak et al., 2013). Stromatolites occur widely in the Precambrian rock record and, because they indicate a history of microorganisms modifying their immediate environment, are accepted as evidence of some of the earliest life on Earth (e.g., Allwood et al., 2006; Schopf et al., 2007). There are many stromatolite and other microbialite discoveries from the early Archean (e.g., Lowe, 1980; Walter et al., 1980; Byerly et al., 1986; Buick et al., 1981, 1995; Westall et al., 2006; Allwood et al., 2009, 2010; Wacey, 2010; Nutman et al., 2016; Djokic et al., 2017; Hickman-Lewis et al., 2018; Homann et al., 2018), a number of which have subsequently been questioned (e.g., Lowe, 1994; Schopf et al., 2006; Bosak et al., 2013; Allwood et al., 2018; Javaux 2019; Lepot, 2020, and references therein). The Assessment Framework (see schematic in Fig. 1) could provide a more universal basis on which the validity of these discoveries can be examined. It can be difficult to assess the biogenicity of any one purported stromatolite, especially if it has been altered through diagenesis and/or metamorphism (e.g., Nutman et al., 2016; Allwood et al., 2018).

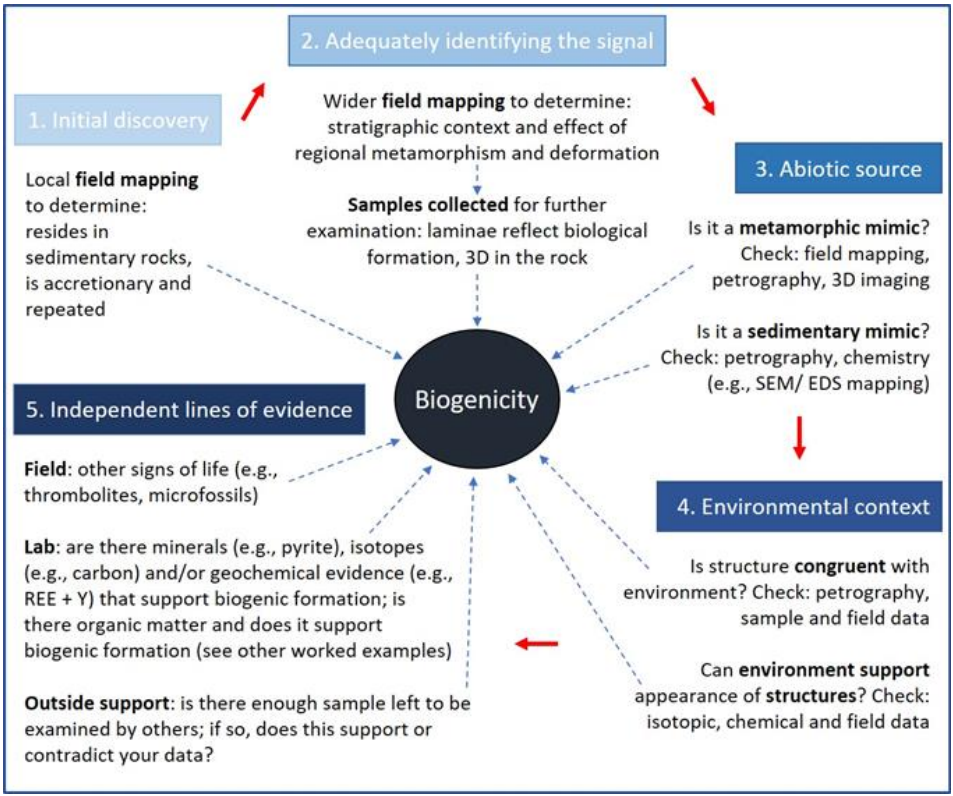

**Figure 3.1**: *Summary diagram highlighting some of the necessary actions (including suggested analyses) required at each step of the framework when using it to assess an in situ detection of an ancient laminated structure. In this context, it is important to recognize that there are multiple lines of evidence at each step in the framework and that each of these contribute to the biogenicity assessment. As you work through the framework, the relative confidence in a detection being biogenic increases (as indicated by the increasing intensity of blue in the headings).*

It is also recognized that abiotic processes can mimic the broad morphology of stromatolites (e.g., Grotzinger & Rothman, 1996; Chaftez & Guidry, 1999; McLoughlin et al., 2008), which can further complicate identification. Though the biological information preserved from early Earth is often patchy and/or overprinted, the biogenicity of a purported stromatolite can usually be distinguished using several lines of evidence across multiple scales (sensu Question 5; Lepot, 2020, and references therein) and by comparison to modern examples and growth models and experiments (e.g., Batchelor et al., 2003; Petroff et al., 2010; Hickman-Lewis et al., 2019).

**Question 1. Have you detected an authentic signal?**

Laminated sedimentary structures of possible biogenic origin are initially identified in the field from visual observations in the field, based on stratigraphic occurrence, bulk mineralogical fluctuations and the macroscopic morphology of the structure and its relationship with surrounding layers (Grey & Awramik, 2020).

**Question 2. Have you adequately identified the signal?**

While in the field, assess whether the laminated structures are primary and indigenous to the rock by observing them from a hierarchy of levels (spanning mega-, macro-, meso-, and microscopic scales), focusing on their various components (e.g., layers, textures, grains, etc.; Grey & Awramik, 2020; Fig. 3.2). Investigate whether they are only present on the surface of the rock (which may be indicative of a mineral deposit or weathering expression), or are embedded in (i.e., are a part of) the rock. Also consider how much the structures may have been impacted or even created by secondary geological processes (e.g., overprinting, veining, mineral replacement, fluid flow). This may require sampling and further analysis in the laboratory (e.g., thin section petrography, electron microscopy, 3D imaging of morphology, geochemical analysis). Replicate field observations and/or analyses may also be required.

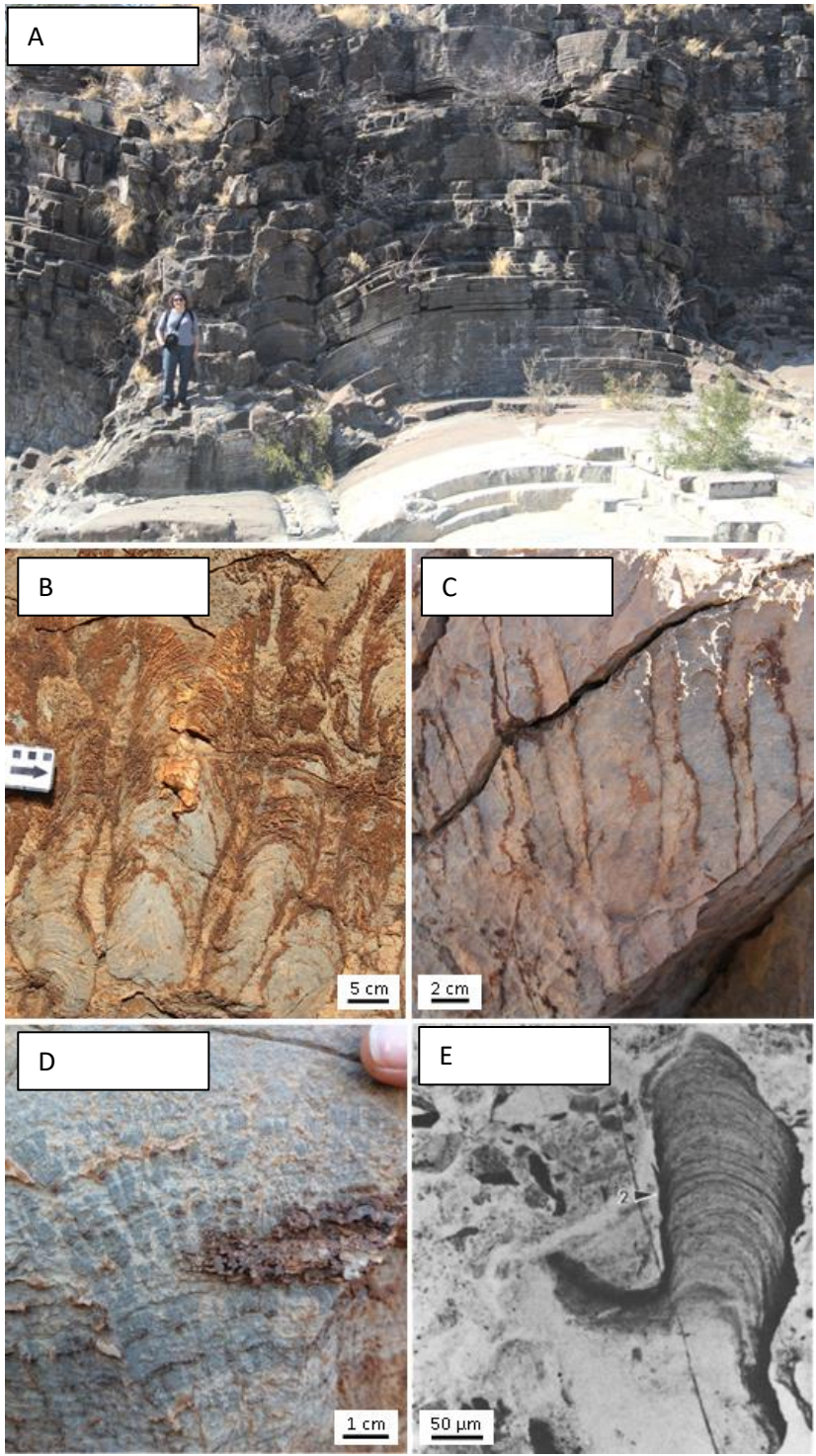

***Figure 3.2***: *Stromatolites can vary dramatically in size, from metres (A), to centimetres (B, C), to millimetres (D), to micrometres (E) in width. This makes it important to assess any laminated structure (i.e., purported stromatolite) at a range of scales. A) Broad, domal stromatolites from c. 2.6 Ga Reivilo Formation, Boetsap, South Africa. B), C) Columnar stromatolites from c. 2.4 Ga Turee Creek Group, Western Australia. D) Microdigitate stromatolites from c. 2.4 Ga Turee Creek Group, Western Australia. E) Micron-sized stromatolite from c. 2.5 Ga Monte Christo Formation, Chuniespoort Group, South Africa. Photo credit: A, AD Czaja; B and D, Barlow et al. (2016); C, EV Barlow; E, Lanier et al. (1986).*

Additionally, one can investigate the interplay between the laminated structure and the surrounding sediments; can you distinguish between the two? In cross section, many stromatolites exhibit vertical relief above the corresponding layer of sediment, reflecting the interaction between processes that are biologically influenced and those that are purely sedimentological/physical (e.g., Van Kranendonk, 2011; Fig. 3.3).

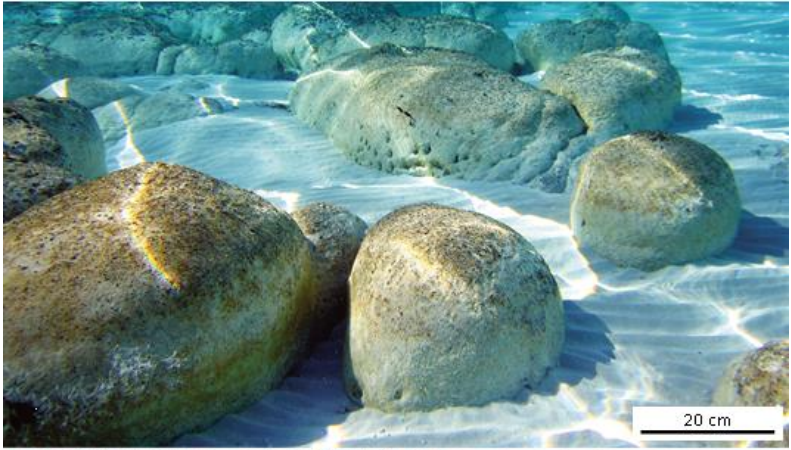

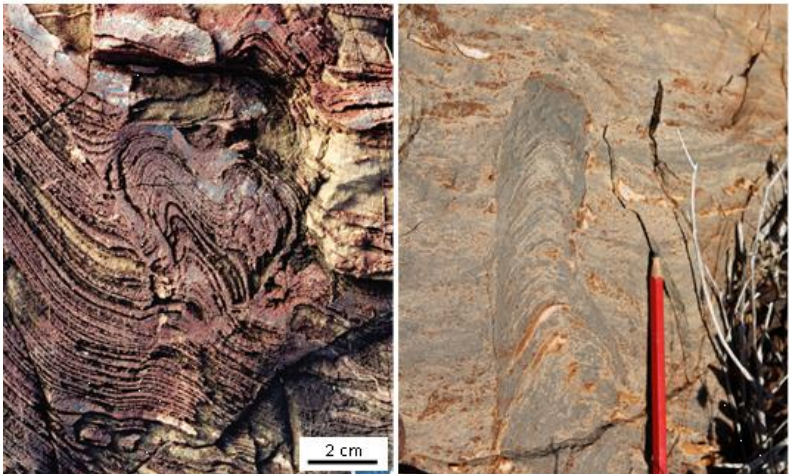

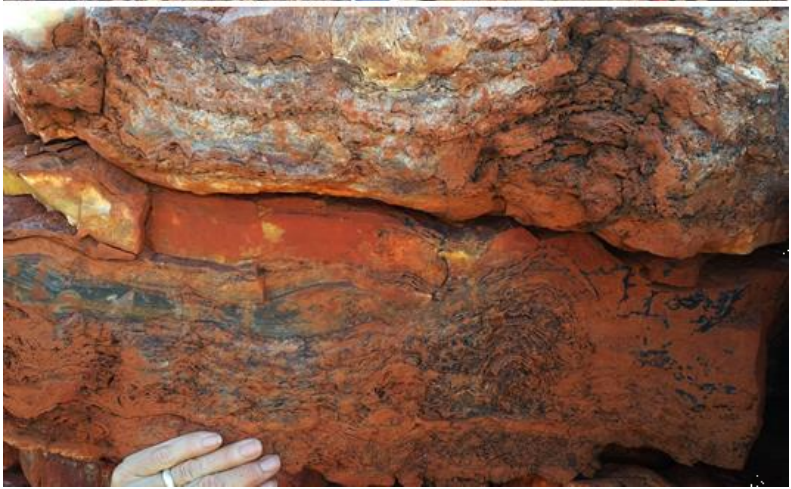

Next, determine whether the laminated structures are both vertically and horizontally extensive throughout the stratigraphy. Stromatolites commonly grow in clusters, or bioherms, with many individual specimens preserved in close proximity to one another (Fig. 3.4; Allwood et al., 2006). If stromatolites are present, then other types of microbialites (e.g., thrombolites, dendrolites, leiolites, or microbially induced sedimentary structures (MISS)) may occur in close proximity, and there may even be a transition from one type of microbialite to another within a single structure (Grey & Awramik, 2020). As such, multiple observations of the same, or similar, type of laminated structure within an outcrop can increase detection confidence. Biogenic stromatolites can also feature micromorphologies incompatible with abiotic morphogenesis. For example: predictable non-isopachous laminations that thicken over the crests of domes due to phototrophic behaviour (Pope et al., 2000); overhangs that maintain shape by virtue of the cohesive properties of microbial sediment (Cuerno et al., 2012); and supra-lamina complexities that develop in response to environmental stressors (Hickman-Lewis et al., 2019).

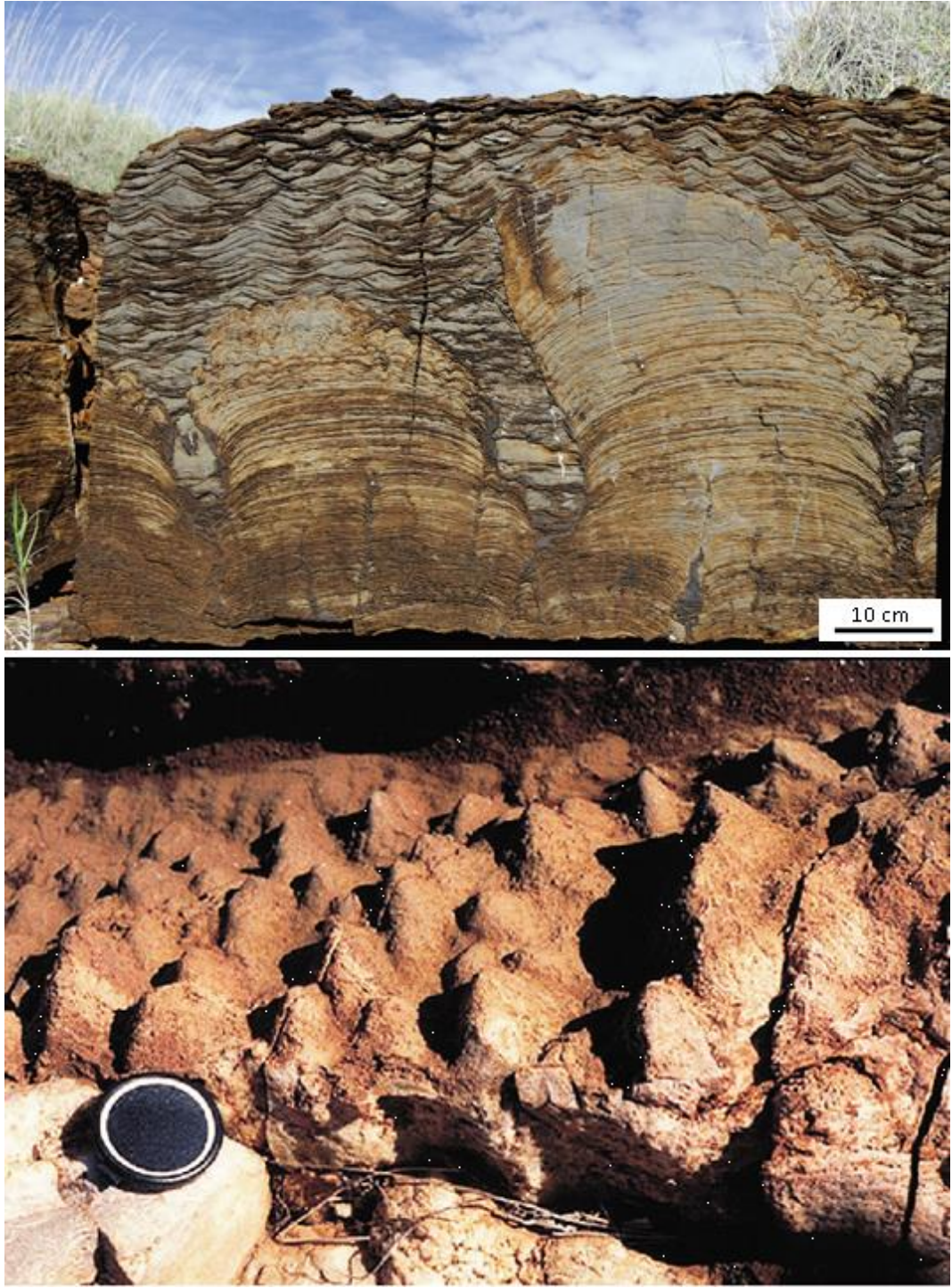

**Figure 3.4**: *Stromatolites often grow in close proximity to one another. A) Large, broad columnar stromatolites from c. 2.7 Ga Meentheena Member, Tumbiana Formation, Western Australia. B) 'Egg-carton' stromatolites from c. 3.43 Ga Trendall Locality, Strelley Pool Formation (sample now housed at the Western Australian Museum); lens cap for scale (6 cm diameter). Photo credit: A, SM Awramik (in Grey & Awramik, 2020); B, Hickman et al. (2011).*

**Question 3. Are there abiotic sources for your detection?**
There may be abiotic sources for detections at each stage of evaluation, from the initial detection in the field, to the search for supporting lines of evidence in the laboratory. For example, metamorphism of sedimentary rocks can mimic some macromorphological features of stromatolites (Allwood et al., 2018; Zawaski et al., 2020). It is necessary to ascertain the deformation history of the unit to assess the likelihood that the laminated structure formed via metamorphism. If the detected structure is in a highly deformed or metamorphosed rock, this might suggest structural deformation is responsible for its formation (Allwood et al., 2018; Zawaski et al., 2020, 2021, although see also: Nutman et al., 2019, 2021).

Abiotic sedimentary processes common to benthic environments may also produce structures that are morphologically similar to stromatolites (e.g., fractal geometry work by Verrecchia, 1996, and Grotzinger & Rothman, 1996; radiating crystal splays in Chafetz & Guidry, 1999; laboratory experiments by McLoughlin et al., 2008). The possibility of abiotic formation of laminae is especially relevant in environments where it may be harder to differentiate the influence of biology on structure formation (e.g., in an evaporative pool within the supratidal zone, or in sinter splash zones in a hot spring system). Careful morphologic, petrographic, and geochemical analyses, coupled with morphogenetic hypothesis building, can help differentiate between biotic and abiotic structures (e.g., Flügel & Munnecke, 2010; Grey & Awramik, 2020).

**Question 4. Is it likely that life would produce this expression in this environment?**
Context is critical in identifying whether a laminated structure may be a stromatolite. Firstly, assess the source of the detection in the context of the regional and local geology. Radiometric tracer studies or geochronological dating techniques could be used to confirm that the sediments were deposited during a time when life is believed to be present. As stromatolites are sedimentary structures, it is essential to verify that the structure formed within a sedimentary deposit. Stromatolites are known from a wide variety of sedimentary settings, including marine, hypersaline, lacustrine, fluvial, hot spring, and cave environments, as well as potentially in soils (Bosak et al., 2013; Grey & Awramik, 2020, and references therein). The detection does not have to be from one of these environments, however, the environment in question should have been habitable for a long enough time to enable the production and preservation of a biological signal.

Stromatolite morphology is strongly influenced by its environment. It is a requisite to assess whether the morphology and/or size of the laminated structure varies, and whether this change is consistent with the physico-chemistry of the paleoenvironment and interpreted water depth (Allwood et al., 2006, 2009; Petroff et al., 2010; Flannery & Walter, 2011; Riding, 2011; Bosak et al., 2013; Grey & Awramik, 2020). By recognizing whether the study site is sedimentary, retains biotic and environmental signals, and is conducive to stromatolite growth, one can determine how likely it is that stromatolites could have occurred in that particular environment.

**Question 5. Are there independent lines of evidence to support a biological (or non-biological) explanation?**
Steps 2-4 outline essential lines of evidence in establishing whether a laminated structure is biological in origin (Fig. 3.4). Other analyses can be used to strengthen and provide confidence in a biogenicity claim. For example, the presence of syngenetic, biogenic organic matter, detected through Raman or other laboratory analyses, can support a biogenic interpretation (see section on organic matter worked example; Pasteris & Wopenka, 2003). Moreover, stable isotope data can support the biogenicity of organic matter and/or minerals within laminae (e.g., Flannery et al., 2016; Marin-Carbonne et al., 2018). The presence of microfossils within the laminae of stromatolites can further support a biological interpretation, although their preservation potential in ancient carbonates, in particular, is considered low (e.g. Buick et al., 1981).

When looking at an individual line of evidence, there may be an abiotic explanation. However, it is unlikely that multiple lines of evidence supporting a biogenic origin will exist in the same sample in the absence of life (Fig. 3.4). Thus, multiple independent and complementary analyses can improve the confidence level of a detection and better assess the validity of a biological explanation.

### 3.2 In situ Worked Example: Detection and Characterization of Organic Deposits on Early Earth and Mars

Understanding the origin and alteration of organic matter on Mars as well as early Earth is of significant importance. Organic molecular structures have the capacity to record substantial amounts of information that could indicate their origins, whether abiotic or biological (see Fig. 3.5). Like early Earth, ancient Martian sedimentary environments exhibit the potential to have preserved ancient crustal organic deposits. A key current area of investigation in this area is selecting the most promising sedimentary Martian samples for collection by the Mars Perseverance rover for future return to state-of-the-art laboratories on Earth.

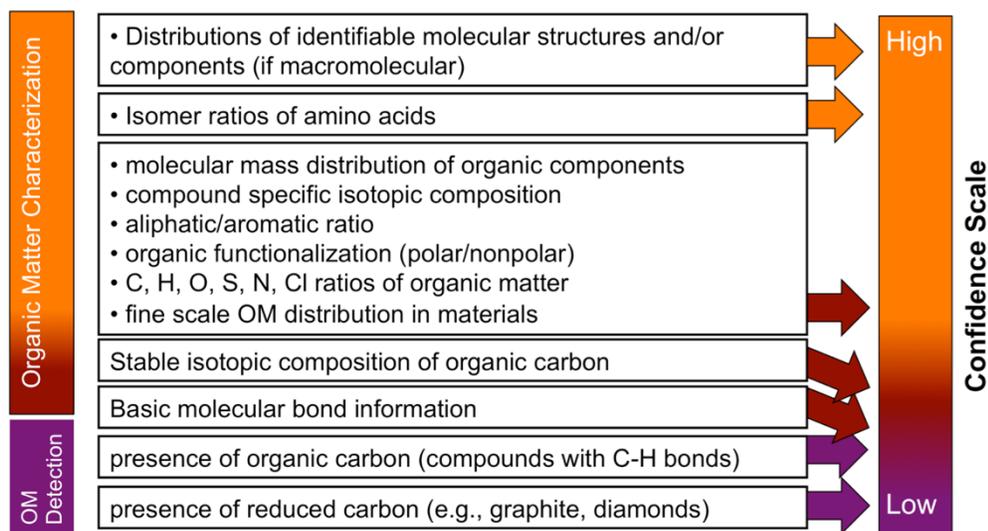

**Figure 3.5** *Organic molecule detection methods are not as definitive as some types of organic matter characterization. Confidence of detecting a definitive biosignature for various observation types for organic matter. Observation types (central column) in order of certainty that the observations could yield a definitive biosignature (right column). Left column denotes general type of observation: detection vs characterization (Adapted from Mustard et al., 2013).*

### Question 1. Have you detected an authentic signal?

The characterization and identification of organic compounds from geologic materials from both Earth and Mars generally rely on mass spectroscopic techniques that can be affected by both sample handling procedures and instrument parameters. For example, the potential for a false positive can arise when tuning is non-optimal, resulting in fragmentation that is ambiguous or difficult to interpret. These erroneous signals differ from contamination since they represent either machine noise or degradation of components that can be mistaken for organic compounds of interest. Careful tuning and maintenance, both in the laboratory and on flight instruments, as well as procedural blanks and standard characterization, can prevent misidentification and indicate unreliable origins of a signal.

## Question 2. Have you adequately identified the signal?

**On Earth**, sedimentary organic matter is the most abundant, widespread, and ancient remnant of our early biosphere. Macromolecular reduced carbon (i.e., 'kerogen') is the most abundant organic component and occurs in rocks as old as 3.46 Ga (Alleon et al., 2019). Solvent-extractable organic molecules that are demonstrably indigenous to their host rocks have been confirmed in deposits as old as 1.6 Ga (Summons et al., 1988). Thermal metamorphism presumably has obliterated these molecules in older deposits, possibly even converting such material to graphite (van Zuilen et al., 2002). Multiple laboratories have utilized multiple techniques (e.g., GC-MS, laser raman and NMR spectroscopy) to corroborate detections of kerogens and solvent-extractable hydrocarbons that occurred principally in shales spanning a range of thermal maturities and representing billions of years in age (e.g., Schopf and Klein, 1992).

**On Mars**, the Sample Analysis at Mars instrument (SAM) on Curiosity rover analyzed multiple portions of the drilled Cumberland mudstone sample in Gale crater and detected several chlorinated hydrocarbons, including chlorobenzene and two- to four-carbon dichloroalkanes, at tens of parts-per-billion (Freissinet et al. 2015). Other ancient Gale crater mudstones yielded a range of other organic compound classes, including thiophenic, aromatic, and aliphatic compounds released from the samples at elevated temperatures >500C (Eigenbrode et al. 2018). Through multiple procedural blank runs and supporting laboratory analog experiments, the SAM team identified an array of contaminants sourced from the instrument, including several compounds resulting from reactions between the samples and a SAM chemical derivatization agent (MTBSTFA) and successfully distinguished them from organic compounds that were indigenous to Mars. Laboratory simulations indicated that the chlorobenzene detected by SAM was derived from reactions during pyrolysis between perchlorate salts and martian benzoic acid in the mudstone (Freissinet et al. 2020). The sources of the organic carbon (i.e., compounds with C-H bonds) are currently unknown and could derive from a variety of Martian (igneous, hydrothermal, atmospheric, or biological) or exogenous sources such as carbon-rich meteorites or interplanetary dust particles.

## Question 3. Are there abiotic sources for your detection?

**On Earth**, reduced carbon constituents in sedimentary rocks are derived ultimately from biological sources; however, some specific kerogen macromolecular structures and solvent-extractable compounds can be created by the chemical alteration and degradation of these constituents by thermal metamorphism, an abiotic process (Sephton & Hazen, 2013). Examples of demonstrably abiotic synthesis of organic compounds are sparse and perhaps limited to low-molecular weight alkanes in the deep continental crust (e.g., Sherwood Lollar et al., 2002) and hydrothermal systems (e.g., Welhan & Craig, 1982; Ménez et al., 2018). Abiotic organic material of meteoritic origin has also been identified concentrated within a specific horizon of the 3.33 Ga Josefsdal Chert (Gourier et al., 2019).

**On Mars**, the organic molecules in the samples analyzed by SAM could have derived from a variety of abiotic sources of either Martian or exogenous origins. It is difficult to assess the likelihood that the organics detected by SAM are abiotic or biotic in origin without additional characterization of the organic molecules, which is not possible with the current instrument suite. Accordingly, the SAM detections of organic molecules on Mars will remain on the low end of the biosignature confidence scale.

**Question 4. Is it likely that life would produce this expression in this environment?**
**On Earth**, sedimentary rocks that harbor the greatest abundances of organic matter were deposited in environments that favored both relatively high rates of primary production and enhanced deposition and preservation of organic matter in clay-rich sediments (e.g., coastal environments, estuaries) (Hedges & Keil, 1995). Given the highly efficient remineralization of organic matter by microorganisms, the abundances of organic matter in sedimentary rocks are largely determined by the efficiency of their preservation in sediments.

**On Mars**, the fine-grained sedimentary rocks (mudstones) analyzed by Curiosity in Yellowknife Bay, Gale crater, were formed in an ancient lake environment that would have been suited to support a Martian biosphere (Grotzinger et al., 2014). The aqueous environment was characterized by neutral pH, low salinity, and variable redox states of both iron and sulfur species. Carbon, hydrogen, oxygen, sulfur, nitrogen, and phosphorus were measured directly as key biogenic elements, alongside a variety of organic molecules detected in the ancient mudstones. Similar lake environments might have been relatively widespread on early Mars.

**Question 5. Are there independent lines of evidence to support a biological (or non-biological) explanation?**
**On Earth**, independent lines of evidence include co-occurring biological structures, biogenic minerals, inorganic metabolites, and stable isotopic patterns that collectively increase confidence that accompanying organic deposits have biological sources (see Fig. 3.5). Multiple techniques could be used such as GC-MS, Raman and NMR spectroscopy, to characterize demonstrably biogenic molecular structures (e.g., hopanes, steranes, porphyrins, carotenoids, etc.) situated in solvent-extractable phases and covalently bound in kerogens.

**On Mars**, more information is needed about the composition and structure of the organic matter detected in the mudstones prior to pyrolysis heating, which can lead to significant modification and destruction of the organics. Chemical analyses of the full distribution of solvent-extractable organic molecules, compound-specific stable C, H, O, and S isotope measurements of organic matter, and spatially resolved measurements that could reveal organic-mineral associations in the samples would all be important in efforts to constrain their origins (abiotic or biotic). These measurements are beyond the capability of the Curiosity payload. Given the SAM instrument's limitations and additional constraints associated with in situ missions (limited time, power, and resources to analyze numerous samples and repeat measurements), the unambiguous detection of an organic biosignature will pose great challenges in situ, and may be impossible. The search for definitive evidence of chemical biosignatures is one of the strongest motivations for Mars Sample Return (Meyer et al., 2021).

### 3.3 Remote Sensing Worked Example: Oxygen Detection on Exoplanets

Diatomic oxygen ($O_2$) is considered a compelling target biosignature molecule in exoplanetary atmospheres because of its strong fundamental absorption bands, clear link to biological production, large atmospheric and geologic sinks, and comparatively small abiotic sources (Owen, 1980; Sagan et al., 1993; Des Marais et al., 2002). However, recent research has identified several possible false positive mechanisms for abiotic production of oxygen on exoplanets, especially those orbiting M dwarf stars (Domagal-Goldman et al., 2014; Tian et al., 2014; Wordsworth & Pierrehumbert, 2014; Harman et al., 2015; Luger & Barnes, 2015; Hu et al., 2020).

Here we consider the application of the biosignature Assessment Framework for the 0.76 um $O_2$-A band on a terrestrial planet orbiting in the habitable zone (Kopparapu et al., 2013; Kasting et al., 2014) of a Sun-like star. Fig. 3.6 shows an outline of the process following the Assessment Framework, and applicable actions are described in more detail under each of the five framework steps below.

### Question 1. Have you detected an authentic signal?

Validation of an $O_2$ detection requires obtaining a sufficient signal-to-noise (SNR) to rule out a spurious signal from an instrumental artifact. A SNR of 5 at a resolution of R=300 for wavelengths less than 5μm, and R=30 for wavelengths > 5μm (Tessenyi et al., 2013) is considered a reasonable threshold for detection of a spectral feature, although a single band detection is more vulnerable to being confused with overlapping features from other species, as well as artifacts due to background, instrumental and processing effects, and stellar systematics. Ideally, potential artifacts should be ruled out via internal vetting and external independent data analysis. The initial interpretation can be strengthened by detections of multiple bands at different wavelengths, comparison of the suite of observations to models, and retrieved abundance estimates. The confidence in single or multiple band detections would also be enhanced by repeat detections over multiple epochs by the same observatory and/or instrument. Confirmation with a different instrument or telescope is likely not immediately feasible for a space-based detection, but could later be pursued to enhance confidence in the detection. Ground-based detections may be more immediately validated via independent observation. Improved modeling and spectral retrieval schemes will support more accurate detection and identification of potential biosignatures.

### Question 2. Have you adequately identified the signal?

The $O_2$-A band absorbs in a region of the electromagnetic spectrum relatively free from other common absorbers, which helps to minimize misidentification (Des Marais et al., 2002; Schwieterman et al., 2018). However, detecting other $O_2$ and $O_3$ bands (see Table 2 in Meadows, 2017; Table 4 in Catling et al., 2018) at adequate SNR will increase confidence in the identification, as the characteristic pattern of $O_2$ absorption is unlikely to be caused by noise or instrumental systematics. Additional $O_2$ bands include the 1.27 um, 1.06 um ($O_2$- $O_2$ collisionally-induced absoprption: CIA), 0.69 um, and 0.63 um bands (Gordon et al., 2017). At very high $O_2$ abundances, $O_2$- $O_2$ CIA may also be seen at 0.445, 0.475, 0.53, 0.57, and 0.63 um (Richard et al., 2012; Karman et al., 2019), though this may be indicative of a false positive scenario (see Step 3). $O_3$ is an expected photochemical byproduct of $O_2$ photochemistry, and detection of ozone ($O_3$) in the UV (0.15-0.35 um), visible (0.45-0.85 um) and mid-infrared (9.6 um) would further support the inferred presence of $O_2$ in the atmosphere (Leger et al., 1993, 2011; Reinhard et al., 2017; Defrère et al., 2018; Olson et al., 2018; Quanz et al., 2018; Line et al., 2019). Potential contamination by telluric $O_2$ lines in ground-based high-resolution observations can be carefully removed via the planet's characteristic orbital velocity behavior (Snellen et al., 2013; Serindag & Snellen, 2019).

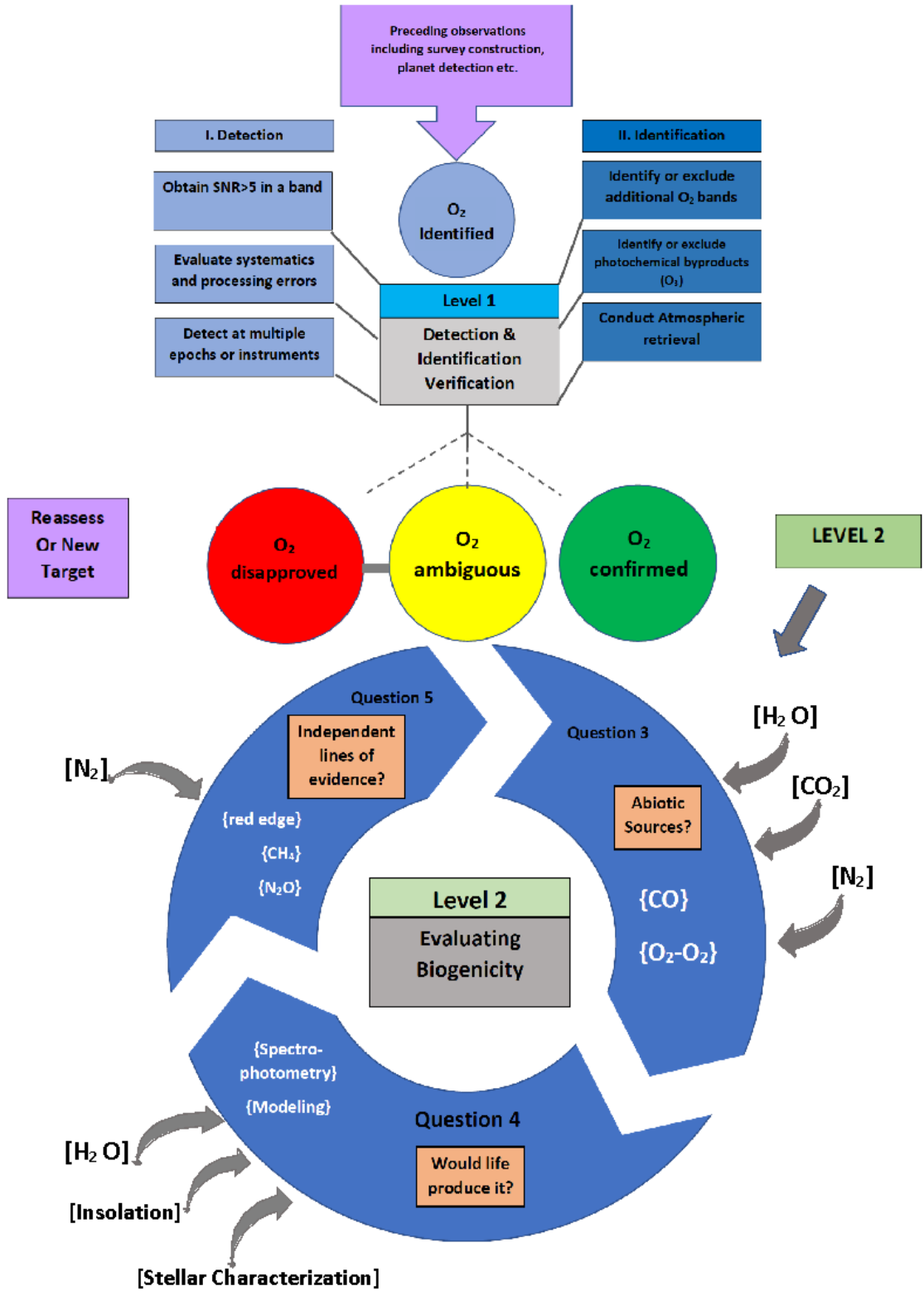

***Figure. 3.6:*** *Schematic process diagram for assessing an initial detection of exoplanetary* $O_2$ *as evidence for a global oxygenic photosynthetic biosphere. Figure Credit: E. Schwieterman and L. Young.*

**Question 3. Are there abiotic sources for your detection?**

Abiotic sources may be distinguished by acquiring additional planetary and astrophysical contextual information. By systematically modeling the larger photo- and geo-chemical environment, we can evaluate how likely it is that the oxygen is produced either by the photolysis of $CO_2$ (Hu et al., 2012, 2020; Tian et al., 2014; Harman et al., 2015) or the photolysis of water and subsequent hydrogen escape (Luger & Barnes, 2015; Wordsworth et al., 2018; Krissansen-Totton et al., 2021). These abiotic mechanisms have also occurred on Mars and Venus. On Mars, atmospheric recombination of photolyzed $CO_2$ is slowed by the lack of water vapor (Nair et al., 1994; Zahnle et al., 2008), allowing this abiotically-generated $O_2$ to build up to 0.1% of the atmosphere, a mechanism that may be even more pronounced on planets orbiting M dwarfs (Gao et al., 2015). Whereas on Venus, the early runaway greenhouse, water photolysis and H loss may have generated an $O_2$-rich atmosphere (Chassefière et al., 2012) that was subsequently sequestered into a magma ocean or the crust, or lost to space (Gillmann et al., 2009; Luger & Barnes, 2015; Schaefer et al., 2016; Wordsworth et al., 2018).

Distinguishing between a true global biosphere and abiotic false positive scenarios may be possible with additional observations of the environmental context (Meadows, 2017; Meadows et al., 2018). These strategies extend to co-detecting additional biosignatures (Seager et al., 2012; Kaltenegger, 2017; Schwieterman et al., 2018) and establishing quantitative, Bayesian frameworks for remote biosignature assessment that help assess whether the signal is more or less likely to be due to life (Catling et al., 2018; Walker et al., 2018). Characterizing the host star's X-ray and ultraviolet radiation, which can vary with stellar age and type of the host star (Loyd et al., 2016; Hidalgo et al., 2018) informs the likelihood of photochemical generation of abiotic $O_2$. For a Sun-like star, many known false positive scenarios become less likely (Meadows, 2017; Meadows et al., 2018). Constraints on the abundance of $O_2$ from the presence or absence of $O_2$-$O_2$ collision-induced absorption (Fig. 3.7; Schwieterman et al., 2016) would indicate whether the planet has a massive post-ocean-loss $O_2$ atmosphere. Quantifying atmospheric $CO_2$, CO and $H_2O$, would help quantify limits on the possible photochemical production of $O_2$ (Fig 3.7; e.g., Schwieterman et al., 2016).

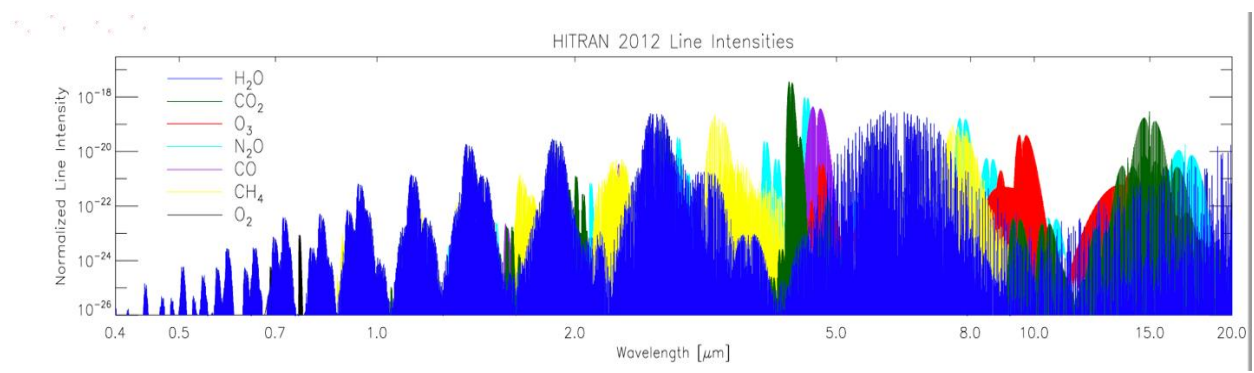

**Figure. 3.7:** *Combined spectral feature catalog for $O_2$-$O_2$, $CO_2$, CO, $H_2O$ and other molecules.*

**Question 4. Is it likely that life would produce this expression in this environment?**

The direct characterization or inference of surface habitability would demonstrate congruence between a putative biosignature detection and the life interpretation (see Robinson & Reinhard, 2020, for a review). If the exoplanet with the potential $O_2$ biosignature is observed within the habitable zone, it will receive stellar insolation that increases the chances for a habitable climate and surface liquid water, but follow-up measurements would be needed to determine whether the planet is, in fact, habitable. Direct detection of a liquid surface could be attempted via glint detection near crescent phase (Robinson et al., 2014), or a combination of surface-type mapping and phase-dependent changes in reflectivity (Lustig-Yaeger et al., 2018). Coupled with other corroborating evidence such as atmospheric water vapor and/or surface temperature constraints this could be used to infer the presence of a surface water ocean. Detection of water vapor coupled with surface temperature and pressure estimates from atmospheric retrievals could also be used to infer habitability (Feng et al., 2018). The evaluation of biosignature congruence with a habitable environment is tied to the environmental assessment for false positives in Step 3 and may require substantial iteration.

Steps could also be taken to determine whether the observed abundance of $O_2$ or other gases are indicative of plausible biological fluxes or build up over time. The availability of metabolic substrates for oxygenic photosynthesis, such as $H_2O$ and $CO_2$, could be quantified, and the history of atmospheric generation or loss of $O_2$ due to stellar evolution could be modeled (Luger & Barnes, 2015; Schaefer et al., 2016) and compared to the measured $O_2$ abundances. Whether biological oxygen accumulation is possible given plausible organic burial rates, oxygen sinks, and photon availability could be modeled (Catling & Claire, 2005; Lehmer et al., 2018). If corroborating evidence for life is found (Assessment Step 5; e.g., coexisting atmospheric $CH_4$), the abundances of the other biosignature gases could be inferred and compared to plausible biological fluxes.

**Question 5. Are there independent lines of evidence to support a biological (or non-biological) explanation?**

Among the most compelling independent lines of evidence to support a biological explanation for exoplanetary $O_2$ would be co-detected biosignatures, including atmospheric disequilibrium (Lovelock, 1975; Sagan et al., 1993; Krissansen-Totton et al., 2016, 2018) through co-detected $CH_4$ or $N_2O$, surface photosynthetic pigments, and seasonal variations in biosignature gases. However, confidence in these secondary biosignatures would also be increased by ruling out their false positives as well. The $O_2/CH_4$ disequilibrium could rule out many false positive scenarios (Step 3) due to the required replenishment fluxes of both species, which varies due to different photochemical destruction rates from different host stars. The 1.65 um $CH_4$ band would be an important target for space-based reflected light spectroscopy, although detecting analog Proterozoic and modern Earth $CH_4$ abundances (< 1.8 ppmv) will likely be challenging (Reinhard et al., 2017; Kawashima & Rugheimer, 2019), but more favorable for Earth-like planets orbiting K dwarf hosts (Arney et al., 2019). $N_2O$ (7.8 and 8.6 um) could also provide additional evidence for a surface biosphere (Rauer et al., 2011) and may be observed with a space-based interferometer (Quanz et al., 2018) or in transit for an appropriate target (Meixner et al., 2019). $N_2$- $O_2$ dominated atmospheres (Stüeken et al., 2016; Lammer et al., 2019; Sproß et al., 2021) especially in the presence of surface liquid water (Krissansen-Totton et al., 2018) may be indicative of biological replenishment. The $N_2$ may be inferred from its photochemical products such as NO and $HNO_3$ (Tabataba-Vakili et al., 2016) or its 4.15 µm $N_2$- $N_2$ CIA feature (Schwieterman et al., 2015). A surface biosphere may be directly detected via surface pigment signatures such as the vegetation red edge (VRE) or analogs thereof (Seager et al., 2005; Kiang et al., 2007; Lehmer et al., 2021) or alternative pigments and reflectance features (Hegde et al., 2015; Schwieterman et al., 2015). Because pigment signatures occur in the same wavelength regime as major $O_2$ bands (see Assessment Step 1), they can be searched for in contemporaneous ultraviolet-visible-near-infrared reflected light observations, although VRE-like signatures will likely be significantly harder to detect than the $O_2$-A band (Brandt & Spiegel, 2014). Seasonal variation in biosignature gases may also be interpreted as additional confirmation of the biogenic nature of the gas (Olson et al., 2018), though quantitatively measuring seasonal changes may be implausible for most scenarios. Ultimately, obtaining independent lines of evidence for the biogenicity of an $O_2$ detection will be highly dependent on the available observing platform(s), as the time between follow-up missions will likely be long. Nonetheless, an $O_2$ detection and confirmation that passes through Assessment Steps 1 through 4 would warrant the additional resources required to search for corroborating signatures. Lastly, biological sources other than photosynthesis (see above) will need to be considered.

### 3.4 An Agnostic Biosignature Worked Example: Distribution of Enantiomeric and Isotopic Composition of Chiral Molecules

The search for evidence of extraterrestrial life in the Solar System has been guided by our understanding of biological processes and, in particular, by direct analogy to specific observations of the chemical and physical expressions of life on Earth. This strategy certainly has merit when exploring nearby terrestrial planets such as Mars, that may share biochemical heritage with life on Earth due to exchange of surface material over billions of years, but becomes less practical when searching for life in the outer Solar System (e.g., Ocean Worlds) where common origins are less likely. The recent Astrobiology Strategy document (NASEM, 2018) specifically recommended that researchers expand efforts to identify and validate biosignatures that are agnostic to life's chemical or molecular framework, particular metabolisms or evolutionary endpoint (Conrad & Nealson, 2001; Cabrol et al., 2016; Johnson et al., 2018; Marshall et al., 2021).

The development of an agnostic biosignature system is a probabilistic process that considers the likelihood that certain chemical and physical structures would arise by abiotic processes by deriving first-principles expectations for a system that relies on energy input unlikely without life. This requires embedded fundamental mechanisms in an interpretive framework rather than analogy and/or anomaly detection. There is an inherent tension in this process, however, since the potential for false positives is greater when the expected signals are too general. Further, an agnostic biosignature framework requires an exquisite prior knowledge of abiotic processes, an area too often overlooked when establishing search parameters. Agnostic biosignatures build from established abiotic baselines and a knowledge of what expressions are not unique to life. This can include potential prebiotic chemistry while assuming nothing about shared heritage. A consistent finding discussed during the Standards of Evidence Workshop pertains directly to agnostic biosignatures - the need for a much better systematic understanding of chemical and geologic expressions in the absence of life (Chan et al., 2019).

A chemical biosignature system with distinct potential as an agnostic concept was described in Glavin et al. (2020). The authors propose a combination of chiral asymmetry, isotopic composition, and molecular distribution as a way to diagnose the biogenicity of a suite of molecules that could be applied to amino acids or polyols (Fig. 3.8). This system builds on the recognition of the overwhelming homochirality expressed in the molecular structures produced by life on Earth. This concept also takes into consideration the fact that racemic mixtures or chiral molecules with much smaller enantiomeric excesses alone may not be a diagnostic feature of life, since similar enrichments due to non-biological processes have been found in meteorites (Glavin et al., 2020). However, if enantiomeric excesses are found in an extraterrestrial sample, a contextual analysis of the isotopic composition of these compounds could be used to indicate if biological fractionation resulting in the isotopic depletion typical of enzymatic reactions has also occurred. A final piece of this system considers the distribution of structural isomers. Biological processes on Earth result in the production of a small number of isomers relative to the enormous diversity of possible compounds of any class, thus a limited distribution could confirm the biogenicity of a suite of chiral molecules. This worked example considers the imagined evaluation of an assemblage of amino acids derived from a geologic sample.

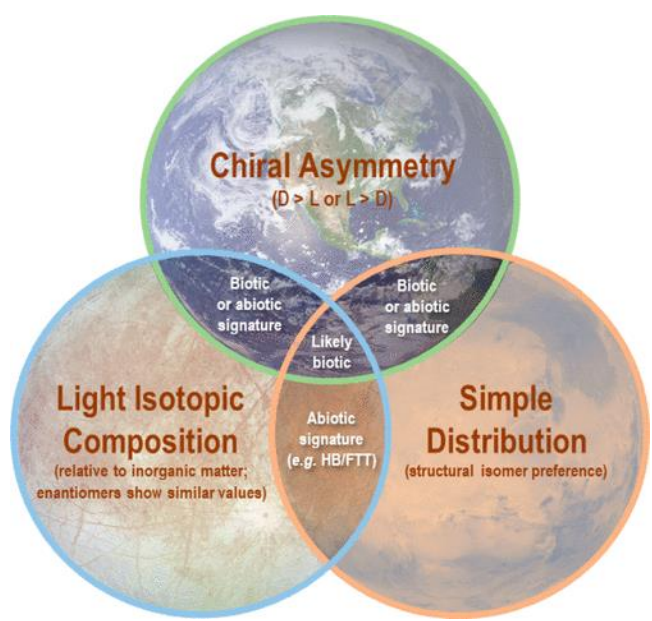

**Figure 3.8.** *Venn diagram showing the three key molecular attributes (chiral asymmetry, simple distribution with structural isomer preference, and a light isotopic composition) needed to establish a likely biotic signature. Note this figure uses the more common D/L notation, rather than the modern R/S absolute configuration as is typical in this field. Images adapted from Glavin et al. (2020) by H. Graham.*

### Question 1. Have you detected an authentic signal?

The techniques necessary to measure both the stereochemistry and isotopic composition of a molecule, and to quantify the relative abundance of molecules, largely rely on different analytical methods, in particular gas and liquid chromatography mass spectrometry (GCMS and LCMS). Both methods also require solvent extraction of the molecules from the sample and chemical derivatization prior to analysis, types of mass spectrometry, etc. As with the early organic example, these methods require a well-maintained and tuned instrument calibrated with standards, a thorough characterization of instrument background, and replicate measurements in order to increase signal to noise and reduce uncertainty associated with molecular identification, enantiomeric ratios and isotopic composition. Compound-specific stable isotopic measurements also require careful standard selection and data correction to avoid misleading fractionation values, since these measurements are reported in reference to an accepted standard (Hayes et al., 1989; Summons et al., 2006; Elsila et al., 2012). An authentic signal could be defined as a measurement above the instrument background (typically above a signal-to-noise ratio of ~3) and would need to meet precision and accuracy metrics that address background ion currents and shot noise (Merritt & Hayes, 1994).

## Question 2. Have you adequately identified the signal?

In order for this system of stereochemistry, isotopic composition and abundance distributions to yield a conclusion with respect to biogenicity, the separation of enantiomers from other potential co-eluting compounds and their identification will need to be highly robust. In the case of amino acids, 96 amino acids have been identified by name in meteorites (including 12 of the 20 most common amino acids found in biology), although hundreds of different amino acids exist in these samples, but have yet to be identified due to a lack of analytical sensitivity and insufficient chromatographic separation for higher molecular weight species (Glavin et al., 2020). A general problem with GCMS and LCMS techniques is also the lack of commercially available standards that are needed to establish the identification of the unexpected or novel compounds we would want to include in an agnostic biosignature framework, particularly higher molecular weight compounds (Glavin et al., 2010; Cooper et al., 2016). If these shortcomings can be statistically evaluated, the systematic differences between the racemic, undepleted, diverse assemblages typical of abiotic organic reactions will be distinctive from biological molecule assemblages that reflect evolutionary selection for constrained chemical properties that perform metabolic functions.

## Question 3. Are there abiotic sources for your detection?

Agnostic biosignatures are the best demonstration of our need for better understanding of abiotic chemical and physical processes. In the case of this example, we can draw on the extensive characterization efforts that have gone into understanding the abiotic organic chemistry of meteorites, as well as recent research describing abiotic organic chemistry in subsurface and submarine hydrothermal systems. When diagnosing the biogenicity of a molecular assemblage by enantiomeric excess, it is important to remember that significant amino acid (and sugar acid) enantiomeric excesses of abiotic origin have been observed in extraterrestrial materials for some compounds (up to 60% L-amino acid excesses and 100% D-sugar acid enrichments; e.g. Cooper & Rios, 2016). In addition, racemization of compounds of either biological or abiotic origins can occur over time due to exposure to liquid water and elevated pressures and temperatures during diagenesis that can completely erase any initial chiral asymmetry (Schroeder et al., 1998; see Fig. 6 from Glavin et al., 2020).

An illustration of several hypothetical relative abundances of enantiomeric proportions in an extraterrestrial sample, and possible origins, is shown in Figure 3.9. Since large L-amino acid excesses of abiotic origin have been found for a few protein and non-protein amino acids in meteorites, the simple detection of a large L-excess in a single amino acid may be insufficient to establish a biological origin. In contrast, the detection of a simple distribution of multiple amino acids that all possess large L-excesses or any amino acids that have large D-excesses would provide more robust evidence for biochemistry, since most amino acids of abiotic origin are racemic, and to date, no significant D-amino acid excesses have been identified in a meteorite or extraterrestrial sample (Glavin et al., 2020). The detection of a D-amino acid excess of any magnitude would therefore provide strong evidence of extraterrestrial origin (since terrestrial biological L-amino acid contamination could be ruled out), and more significantly, of a potential second genesis of life in our solar system. The most challenging scenario would be the detection of a racemic mixture of amino acids as these compounds could be abiotic in origin or the result of complete racemization of an extinct biota. In this case, isotopic measurements could be critical to confirm a possible biological signal, but this requires an understanding of the isotopic composition of source reservoirs and not just a comparison to typical Terran metabolic values. Since racemization does not lead to measurable isotopic fractionation of amino acid enantiomers and chiral amino acids of abiotic origin in meteorites are all enriched in $^{13}$C, $^{15}$N and D relative to biological amino acids (Glavin et al. 2020), the detection of highly depleted isotopic values in both enantiomers of a racemic amino acid mixture could provide strong evidence of an extinct biological source with the caveats mentioned previously. The best correlation to support a biological origin for an assemblage of amino acids would be a simple distribution of chiral amino acids with large enantiomeric excesses (either L or D) and with highly depleted isotopic values relative to inorganic matter in the sample. However, like enantiomeric excess, simple distributions of straight-chained amino acids with light carbon isotopic composition of these compounds can then be used to confirm a possible biological signal but requires an understanding of the isotopic composition of source reservoirs and not just a comparison to typical Terran metabolic values. The best correlation to support a biological explanation for an assemblage would be a depleted isotopic value of the more abundant enantiomer. Like enantiomeric excess, simple distributions can also be produced by mineral-catalyzed or hydrothermal reactions (Islam et al., 2001; Burton et al., 2012); however, these conditions generally produce achiral products. Hence, the most reliable interpretations will rely on the combination of all three measurements.

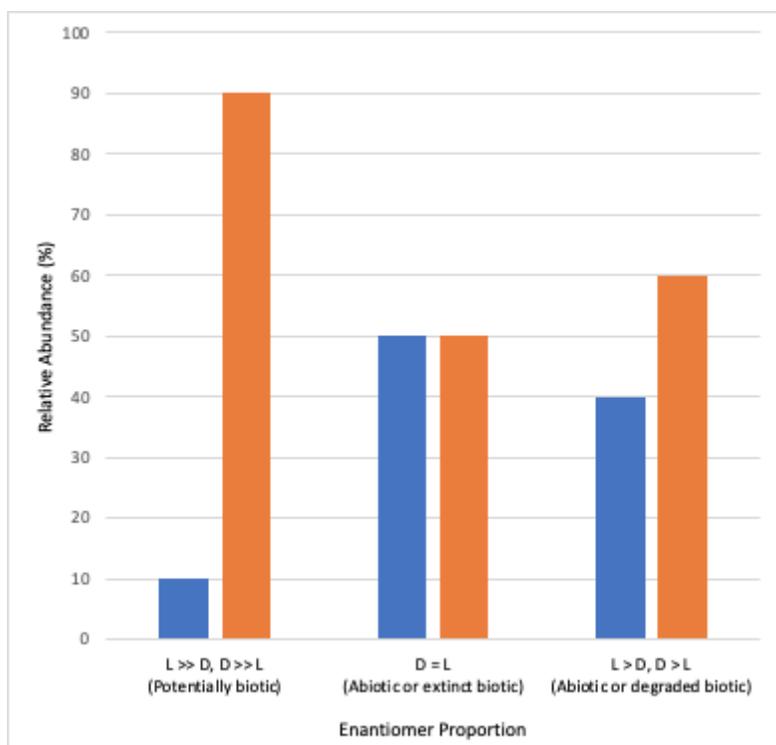

**Figure 9.** *Some hypothetical relative abundances of D- and L-amino acid enantiomers and their possible biotic or abiotic origins (Adapted from Glavin et al., 2020).*

**Question 4. Is it likely that life would produce this expression in this environment?**
Diagnosing the likelihood that compounds in a molecular assemblage would derive from a living system could be accomplished by considering the thermodynamic probability that the physico-chemical formation conditions would produce the same structures with similar stereochemical features. This observation would not require prior knowledge of the metabolism or chemistry of potential biota. Further, knowledge of the physical conditions could indicate if preservation of these molecules is even likely or if an exogenous source is needed to achieve the observed molecular abundances.

**Question 5. Are there independent lines of evidence to support a biological (or non-biological) explanation?**
Given the numerous abiotic reactions that could produce these patterns, confirmation should include spatially resolved chemical, mineralogical, and isotopic studies to establish the origin of organic assemblages. Incubations or analog experiments could be performed to understand novel formation mechanisms or to see if there are physical parameters that promote production and can indicate either a chemical or a biological pathway.

## *4. Applications of the Framework*

### 4.1 Utility of the Framework
Implementation of the Assessment Framework in mission designs starts with defining the science objectives, investigative strategy, concept of operations, requirements flow, and hardware build, followed by mission science return reporting and iterative data interpretation. Starting from conceptualization through operations and decommissioning; this Framework will be useful throughout the lifecycle of biosignature detection missions. Observatories that are not designed for life detection may benefit from this Framework to initiate such a search. Mitigating sources of contamination and establishing biogenicity are of paramount importance and the Framework can be a powerful tool. The Framework can also be used to anticipate experimental, methodological, and statistical needs that build in estimates of uncertainty for observations and interpretations.

### 4.2 Mission-Sensitive Application of the Framework
Confirmation of a detected signal (Question 1) begins with understanding calibration, background noise, sensitivities, resolution, operations, and behavior of instruments under unforgiving environmental pressures of space. Establishing confidence in the signal (Question 2) requires a solid understanding of sources that could impact data interpretation. Pre-flight analog testing and stringent contamination control and knowledge procedures will help distinguish data artifacts and false positives. Confirmation of a detection during a mission can also be supported by multiplicate measurements with multiple instruments investigating the same target body. The ubiquitous presence of Earth's biosphere has limited our knowledge of the natural expression of abiotic physical and chemical systems, however, false positives (Step 3) can be recognized by understanding measurement capabilities for known abiotic signals in a given environment. Question 4 requires that the Science Traceability Matrix (STM) connects measurements back to key aspects of habitability and the context of the planetary system in order to determine if the biosignature in question would likely be produced in this system. The STM should also assess sample concentration and preservation. Similarly, Question 5 requires that the renewed STM trace multiple measurement techniques to determine if independent lines of evidence support biological (or abiotic) formation. The value of independent lines of evidence, with detectable interdependence, consistently pointing in the same direction strengthens confidence in the overall conclusion.

## 4.3 Mission Life Cycle Considerations

Implementation of the Assessment Framework could prove useful throughout the lifecycle of missions, from conceptualization through operations and decommissioning.

During pre-Phase A, a mission concept is developed, often starting with a team that studies the mission concept, with the objective of understanding how difficult it would be to carry out the mission's science goals. The output from those teams outlines the science goals, defines options for mission architectures and instruments that could realize those goals, and prioritizes the technologies that would be needed to enable or enhance those architectures and instruments. NASA-led studies of this type are usually referred to as a "Science Definition Team" for planetary missions and as a "Science and Technology Definition Team" for astrophysics missions. During this phase of a project, the Framework would help teams determine which, if any, levels of the Framework would be part of the mission concept's goals and allow the teams' guidance on what specific questions and investigations would be required to meet those goals. The Framework could also help formalize and motivate decisions made during these studies. In particular, the mission may benefit from a biosignature verification plan---potentially outlined at the proposal stage---that would consider the measurements needed, and potential precursor or parallel work needed to support the mission's biosignature detection and assessment capabilities.

As an example, consider the final reports from the LUVOIR and HabEx mission concept teams. Although the Framework did not exist when those studies were conducted, both teams included a search for biosignatures and studied concepts that would meet at least the first four stages of the Framework. Both teams studied architectures with wavelength ranges that would allow for the detection of biosignatures (Step 1) evident either on modern-day Earth (e.g., $O_2/O_3$) or thought to have been evident in earlier eons of Earth's history (e.g., $CH_4$ or an organic haze). They also explicitly ensured "contamination" (Step 2) would be limited by studying the potential for background objects to confuse their detections and by considering observing strategies that would eliminate the potential for light from the Earth or Moon to enter the instrument array. Both teams considered abiotic production mechanisms (Step 3) for the biosignatures they were designed to detect and interpretive strategies to identify these abiotic mechanisms. These strategies included both a wavelength range on the "biosignature detection" instruments that would add environmental context to the detection, and requirements for a second instrument that would measure the UV flux from the star, which is an important energy source for abiotic production of atmospheric biosignatures. These studies made sure the instruments responsible for detecting these molecules had the required spatial resolution. Similarly, both missions included a separate set of observations to confirm the presence of liquid water on targets. With its larger aperture, LUVOIR is also capable of searching for biosignatures associated with pigments or surface vegetation and could begin addressing Question 4 of this Framework. LUVOIR and HabEx observations could complement one another (Question 5) if an ideal target becomes available. All of this - the detection of the biogenic molecules, avoidance of contaminant light sources, discrimination of false positives, and confirmation of global-scale surface habitability - were explicitly included in the teams' Science Traceability Matrices, at the level of top-level science goals that flowed down to specific instrument properties.

These instrument properties were in turn used to help determine the technologies that would enable or enhance the missions and ensure they could collect the requisite data. These technologies were subsequently available for funding through NASA's competitive Technology Development for Exoplanet Missions (TDEM) call. The totality of this was presented in the final reports available to the community. Additionally, this allowed engineering teams - led out of NASA GSFC in the case of LUVOIR and out of JPL in the case of HabEx - to provide details on the architectures and instruments that would achieve the top-level science goals. This enabled multiple outside groups to create "ballpark estimates" for the complexity and associated costs of these projects, so that these ambitious science goals and their associated costs could both be considered as part of the Astro2020 Decadal Survey.

In later phases of a mission, the Framework will be similarly useful. The Framework can help determine the necessary technologies and help prioritize the funding of those development activities. Similarly, top-level mission architecture decisions can be made with specific Framework goals in mind, including incorporating Framework goals as primary mission goals with requirements for selected instrumentation to meet. Such requirements are used as benchmarks for performance for the mission and its instrumentation throughout the design, fabrication, assembly, integration, and testing phases. Placing specific Framework steps in the mission's top-level science goals is the best way for an interdisciplinary team of engineers and scientists to design a mission that could achieve the life detection goals laid out and identified by a concept study team.

During operations, the relationship between the Framework and the mission comes back to the concept of confidence – a measure of certainty on what has (or has not) been detected. This would dramatically impact mission operations planning, the publications from the mission, and the communication of results from the mission. For an example of how this would work, consider a rover on the surface of Mars. To achieve Questions 1 and 2 of the Framework, a rover would need to have the ability to detect a signal from life while ensuring decontamination steps, calibrations, and blank analyses - all of which are necessary to demonstrate the status of the instrument signal, noise, and operations, as well as ensure that there are no forward contaminants that might compromise analyses. Addressing Questions 3 and 4 of the Framework would likely require a coordinated use of instrumentation. Confirmation of a detection could also be supported by other measurements of the same feature by the same instrument, other instruments on the mission, and/or instruments of other missions. The specific approach will depend on the nature of the possible biosignatures being investigated on a mission, and the known likelihood that they may also be products of abiotic formation processes. That said, all biosignature investigations will benefit from contextual measurements that support characterizing both the sample, as it may help understand other instrument results, and details of the potentially habitable or inhabited environment. In applying Question 4 of the Framework, the mission should trace selected contextual measurements back to key aspects of habitability that may have the greatest value in assessing whether life would produce the biosignatures being investigated. Step 5 of the Framework applied to *in situ* missions requires that the STM traces a biogenicity investigation to multiple measurement techniques to determine if independent lines of evidence support biological or abiotic formation. As on Earth, it is expected that extraterrestrial life would impart many signatures that are independently detectable but exist interdependently. Thus, the value of independent lines of evidence (e.g., lipid patterns, elemental distributions, and microbial mat structures) supporting a consistent interpretation instills much more confidence in the overall conclusion compared to any single line of evidence. Further, any detectable interdependency among the biosignatures may provide yet another orthogonal interpretative path to assess confidence (e.g., distribution of amino acid skewed to complex structures, enantiomeric excess, and peptide chains; or cellular structures and/or motility).

Application of the Framework to *in situ* biosignature-seeking missions should be continuous throughout the life-cycle of a mission, which may be decades. Following mission design, decisions for modifications and refinements to the build (e.g., instruments and contamination control) and implementation plans (e.g., resources for cruise and science operation activities, concept of operations) need to consider the impact of changes on how well the mission maintains the capability to address the Framework. The Framework provides high-level guidance that if implemented throughout the mission process (i.e., concept, design, build, launch, cruise, science operations, data analyses, and reporting) enables mission scientists to establish a knowledge base from which to assess and understand confidence assigned to biological (or abiotic) interpretations.

## 4.4 Confidence and Statistical Inference

Drawing on lessons learned from previous life detection claims that, on reevaluation, rested on faulty statistical reasoning; attendees at the Standards of Evidence workshop agreed that researchers should collaborate with statistics specialists when reporting statistical claims. Statistical methodology is critical in both the detection (Questions 1 and 2) and assessment (Questions 2, 3 and 5) levels of the Assessment Framework.

### 4.4.1 Biosignature Detection

No experimental result or claim is complete without an attendant measure of its uncertainty. This is particularly important for biosignatures since sample sizes are likely to be small, evidence marginal, and impact enormous. As such, care should be taken to follow standard methods to report uncertainty clearly and to increase the reproducibility of claims (Rein, 2019). All experimental results contain uncertainty resulting from resolution limitations and instrument and methodological systematic error, as well as random process uncertainty due to finite sample effects and instrument precision, and uncertainty in inductive and statistical inference methods.

Scientific conclusions are reached through deductive and inductive reasoning, with induction being required to make conclusions from empirical evidence. Given that biosignatures are empirical in nature and, by definition, induction cannot prove a conclusion but only lend strength to competing hypotheses, all biosignature detection reports will be uncertain. Methods exist to quantify the contribution of each of these sources of uncertainty, and to measure the interaction and propagation of these sources of uncertainty when combined by multiple layers of analysis.

Though there are well established inductive inference rules, such as the methods of residues, that will undoubtedly be included in any report of biosignature detection, statistical inference is the most widely used framework for quantifying inductive uncertainty due to random processes. The uncertainty here is the result of the limited sample sizes available to experimentation causing the measured values to diverge from the true or population values. Statistical methods are powerful but require specialized techniques for understanding the assumptions being made by the methods precisely (e.g. the number of ties present in two populations of values is critical to interpretation of a Wilcox test). Statistical methods can only be used to quantify uncertainty due to random processes, they cannot be used to quantify uncertainty due to systematic processes, or even detect the presence of systematic error. Drawing on lessons learned from previous life detection claims that, on reevaluation, rested on faulty statistical reasoning; attendees at the Standards of Evidence workshop agreed that researchers should collaborate with statistics specialists when reporting statistical claims.

### 4.4.2 Biosignature Assessment

Depending on whether a Frequentist, Bayesian, or Likelihood approach is being used, uncertainty quantification will usually be in the form, respectively, of p-values and confidence intervals; posterior distributions and their credibility intervals; or likelihoods. Researchers should be conscious of the pitfalls in generating p-values in particular, such as the large sample effect. P-values should be presented in conjunction with the attendant effect size (Sullivan and Feinn, 2012). Confidence and credibility intervals provide a more robust method of quantifying uncertainty (though they depend on the underlying statistical methods being properly applied). Common pitfalls include multiple hypothesis testing without p-value correction (the Bonferroni correction is the simplest and most commonly applied method, though not the only method available) and overfitting polynomials to data or fitting a model to too few data points (i.e., the number of measurements should not be less than the number of fitted parameters). Multiple-hypothesis correction guards against post-hoc hypothesis testing, in which hypotheses are selected on the basis of their statistical significance (also known as p-hacking), and HARKing (Hypothesizing After the Results are Known), in which hypotheses are generated for testing after the experiments have been conducted (Bishop,1990).

An essential component dictating confidence is the Bayesian likelihood of a given signal being produced abiotically, p(signal | no life). In many cases this factor is computationally intractable, and highly vulnerable to model assumptions. Correspondingly, historically many abiotic false positives were unrecognized at the time of original publication (oxygenated atmosphere false positives, Meadows, 2017; abiotic microorganism-like features; McKay et al., 1996; isotopic and mineralogical signatures; Alleon et al., 2019; etc). Using the Bayesian framework helps to mitigate uncertainties in individual target interpretation by addressing population-level questions, such as the likelihood that a survey has found life on at least one target, and makes inferences more robust.

A Bayesian statistical methodology has the advantage that assumptions and prior-information can be encoded into a prior distribution, allowing quantification of uncertainty to be driven by observation. Frequentist methods are equally powerful, but prior beliefs are encoded in the choice of statistical method. For example, a researcher who assumes a non-parametric population distribution will likely choose a Wilcoxian test over the t-test. Bayesian methods would include that assumption in the prior distribution. Whatever the statistical methods employed, the assumptions being made should be clearly stated so that the scientific community can evaluate the merits of those assumptions. Statistical methods should be used to strengthen inductive arguments and Bayesian, Frequentist, and other schools of analysis should reach the same conclusions when properly applied - except when data is so limited that statistical power is severely limited (Albers et al., 2018). When combining p-values from multiple independent variables in a meta-analysis to provide support to a particular hypothesis, Fisher's combined probability test should be considered.

Systematic uncertainty results from consistent error in an instrument or methodology. Since systematic uncertainty cannot be detected through statistical methods, it is critical that the experimental design be such that it is robust to systematic error. However, the potential impact of systematic error can be crafted in tandem with instrument design. Sensitivity analysis before and after experimentation can bound the impact of systematic error and guide experimental design to mitigate it. That is, analysis should be performed that relates the magnitude of systematic error in an instrument to the ability of the experiment and data analysis procedures to differentiate hypotheses. Unlike uncertainty due to random processes, systematic uncertainty can be eliminated entirely by careful experimental setup or by *post-facto* correction.

Two commonly used uncertainty propagation methods are worst-case and differential error quantification. Worst-case analysis measures the maximum error that could be accumulated at each step - from instrument resolution and precision error. These bounds are determined by instrument calibration tests through confidence or credibility intervals for a multi-stage analysis and report the maximum uncertainty that could have accumulated using the additive rules of uncertainty, with appropriate unit conversions. When performing linear regression on data that includes measurement uncertainty, the standard way to retain information about uncertainty in the best-fit slope is to calculate the ratio of the best fit slope to the slope given by the worst-fit, where the worst-fit line is bounded by the measurement error bars. Farrance & Frenkle (2012) provide a review including a summary of terminology to be used in reporting and discussing measurement uncertainty. Biosignature scientists should also be familiar with the "Guide to the Expression of Uncertainty in Measurement" detailed in ISO 15189. By communicating that uncertainty through the confidence or credibility intervals resulting from appropriate statistical tests a researcher can place their discovery on a spectrum of confidence and provide a measure of certainty.

## 5. Using The Assessment Framework For Life Detection Communications

While the main workshop goal was the development and adoption of a generalized biosignature Assessment Framework, an afternoon of discussion was reserved for considering areas and activities where the Assessment Framework could be useful in communicating results. This activity was intended to generate topics that would be continued in a future workshop focused on designing "best practice" reporting protocols that guide biosignature discovery public announcements. Participants agreed ideas generated during this time would remain preliminary since this would clearly need to include stakeholders in the publishing, journalism, and communications fields. Given that the vast majority of workshop participants were scientific researchers, reporting protocol suggestions discussed at the workshop were considered to be "for scientists, by scientists" and focused on actions that we in the scientific community could consider to improve our communication within the community.

### 5.1 Developing Reporting Protocols for Life Detection Claims

Communicating evidence in the search for life to decision-makers, the wider scientific community, and the general public requires consideration of the means by which each audience obtains and interprets information. Given the great importance and high-profile of life detection announcements, it is vitally important that the community facilitate clear communication and incentivize reporting practices that offer correct, concise information on the nature of any detection,  resisting overstatements and correcting any misinformation. These workshop discussions draw on lessons learned from previous scientific efforts in other fields that require the communication of complex information with wide public interest to the greater community.

### 5.1.1 Ideal goals for community biosignature verification and reporting

The first activity in this discussion was for participants to outline ideal goals for a biosignature verification and reporting protocol that they 1) welcome seeing their colleagues and  the greater scientific community follow but also 2) would not find difficult to follow as well. The ideal goals would ensure that both the scope and level of certainty in reported results are appropriately perceived by both the scientific community as well as the public.

*Verifying results and interpretation*

A team reporting a life detection claim would ideally be large enough to accommodate the diversity of discipline specialists necessary to provide a thorough validation of the methodology and results. While the focused research areas that address questions of astrobiological interest may be small, these life detection research activities span a broad variety of methods, instruments, and specializations that may have independent experts outside of the community. Independent verification should be encouraged as a high priority for high-profile results, either pre- or post-publication. Mechanisms to facilitate the voluntary sharing of data and results prior to publication should be explored as an important component of the verification phase of a life detection announcement. Workshop discussions did recognize that there is an inherent tension between the speed of sharing and the rigor of peer review, as well as a clear danger that pre-print materials as they currently exist can be over-interpreted as final results. Participants agreed that new mechanisms may be necessary to facilitate this information dissemination. Community collaborative networks (e.g., NExSS and NFoLD) could serve as nodes to coordinate information sharing and verification research activities and also to provide an additional forum to discuss results in addition to or before the peer review process.

*Reporting to the scientific community*

Discussants suggested a number of possible modifications to the typical publication timeline that would incentivize the verification process for life detection claims. One idea was the addition of a "two-factor" publishing process that couples discovery and verification. Participants also considered a system where biosignature discovery claims and verification/assessment activities could be reported in a stepwise manner, with reference to the stage in the assessment framework for context, and discrete confidence estimates included at each step. This would allow the community to respond to individual interpretations rather than the nested assumptions of an aggregate study. This could also prevent misunderstanding based on in-group language use. Claims should be carefully reported with contextual knowledge (Question 4) to communicate the scope of the discovery. High-profile missions could potentially promote constructive criticism by engaging the broader scientific community in constructing alternate data interpretations.

*Reporting to the public*

In keeping with a general theme that biosignature science should rely more heavily on discipline specialists, workshop participants found that clearly communicating where we are in the scientific process of life detection claims would benefit from more collaboration with science communication specialists. It was agreed that there should be an effort to communicate the nature and details of the Assessment Framework to key sources such as institutional press officers and respected science journalists. Participants discussed reasons to discourage journalists, communications specialists, and even our colleagues from relying on pre-print, pre-reviewed results. This, again, would require prior coordination, possibly with a network of trusted science communicators. Workshop participants agreed that claims directed to the public should clearly communicate that discoveries are usually at *the beginning*, not *the end*, of a lengthy scientific process. That process will likely require subsequent measurements or even missions, which may provide data that result in modifications of the current interpretations.

### 5.1.2 Current obstacles to ideal reporting protocol

In addition to envisioning ideal goals, participants were encouraged to consider the roadblocks that would prevent adoption of these verification and reporting procedures. In the current academic climate, there is immense pressure to publish as many papers and as rapidly as possible. This is particularly true when high profile results are considered. Given that these activities are directly tied to measures of job performance, employment opportunities and/or grant success, an incentive program that encourages researchers to follow the protocol would require a community-wide change in how we assess productivity. Funding agencies would be  necessary partners in helping institutions re-imagine how more rigorous, incremental science could become part of high-profile endeavors. The proliferation of private investment into academic research activities may also affect the motivation to follow the protocol. In these cases, it may be beneficial if community organizations (e.g., NExSS, NfoLD, NOW, ExoPAG, MEPAG, OPAG) do more to invite these researchers to participate in the protocol measures and share insights on how to implement the Assessment Framework.

One of the more challenging elements of a lengthy verification process is how to maintain a discovery embargo when large groups of scientists are involved. Workshop participants discussed the strict non-disclosure agreements within a discovery and verification collaboration that would be required. Further, the concern of "getting scooped" is a powerful disincentive and encourages rapid publication without data sharing for independent verification (see Fig 5.1).

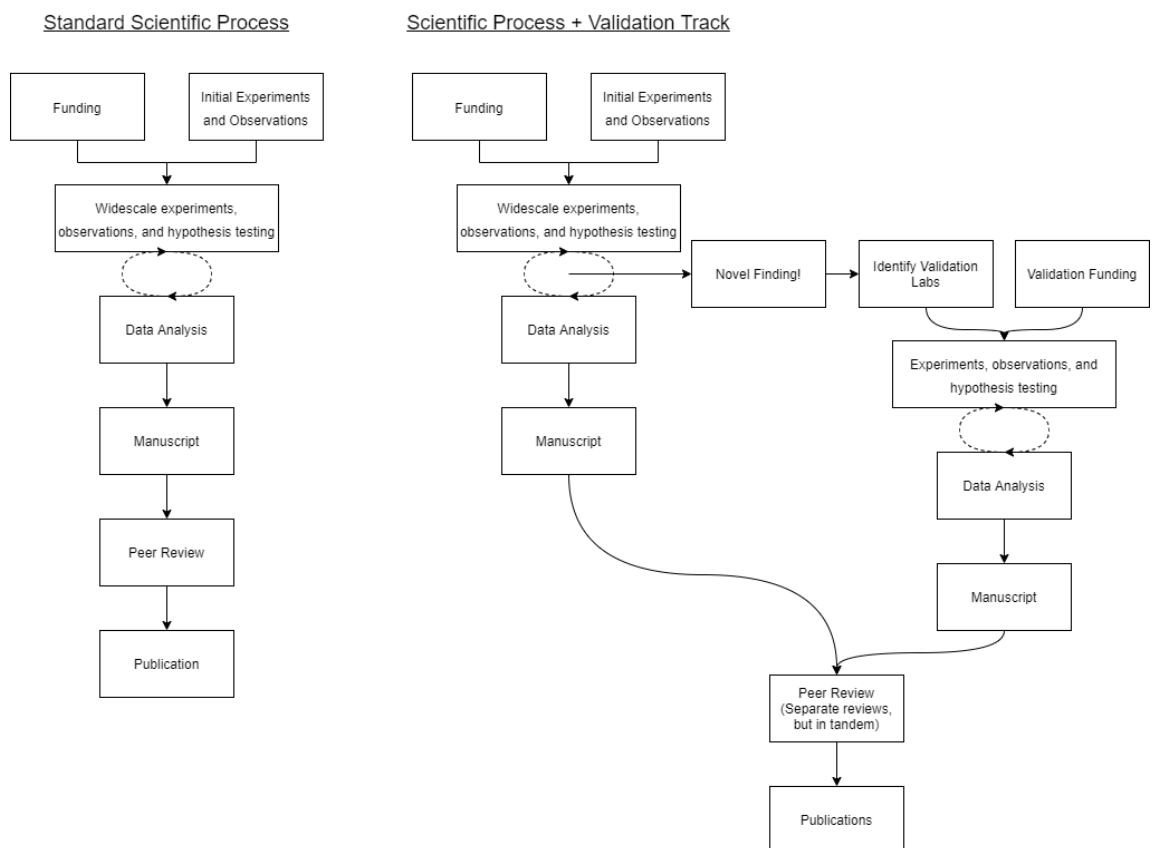

***Figure 5.1***. *One potential publication workflow for a scientific discovery incorporating outside validation. In the left panel, the standard scientific process that leads to publication follows a predictable pathway.*

*The right panel illustrates a new "validation track" workshop participants discussed whereby with initial pools of traditional funding, researchers will progress and iterate with increased data, leading to the testing of more precise hypotheses.*

Workshop participants agree that a community-led culture change is necessary to ensure that individuals are rewarded for their roles in a discovery in a way that *also* incentivizes a robust verification process. Recognition beyond the scientific community can be tied to career advancement and is a factor that could work against widespread adoption of a verification protocol. Individual institutions may also reward controversial results published in high-impact journals that increase an institution's profile rather than building incentives encouraging a robust and iterative scientific process. Without strong community support there will be those who choose not to put effort into progressing through the protocol, or who ignore it for their own benefit. While there is no way to prevent this behavior, effective communication to researchers and institutions about the Assessment Framework and the benefits of these reporting protocols will be needed.

A final obstacle specific to large collaborations are the multiple, possibly conflicting sets of regulations by government agencies, national laboratories, and international collaborators. For example, it may be hard for U.S. civil servants such as NASA employees to participate in oversight activities. Additionally, all stakeholders would need to agree to a reporting protocol and agree on how these processes would guide discovery announcements.

### 5.1.3 Incentivizing the best possible verification and reporting process

For high profile science results, the community should foster an environment where the preference and priority is given to maximizing robust scientific processes. However, except for possible publication, there is currently no direct reward or benefit for groups who verify results. A potential solution to this is a discretionary life detection verification fund — national agencies could have funds available for verification expenses, and acknowledge it as a funding source. Via mutual agreement the verifiers could be included as co-authors on the discovery paper, or publish a companion verification publication, given the significant work likely needed to verify a discovery. This embedded verification progress would encourage greater confidence in findings. Verification experiments often come at a cost that researchers must forward fund, especially to perform them with a desired rapid response time. A responsive funding process would make it possible to build a verification process that does not significantly delay progress, and aims to reduce the time between claim and re-analysis, thus improving data and transparency. Engaging with the Assessment Framework and reporting protocols should be seen as a way to strengthen the credibility of bold claims and not as a regulatory hurdle that penalizes discovery.

Changes to the incentives for reporting life detection to the scientific community must be centralized at the journal level. Too often journals do not accommodate or publicize the caveats to a discovery with the same enthusiasm as an extraordinary claim. Workshop attendees also suggested a more broad adoption of the practice of publishing reviewer reports along with papers as a way to broaden scientific discourse. Another mechanism discussed was a new type of publication that can serve as an interim report describing the progress an observation has made along the Assessment Framework and the steps that have been taken in an investigation. This would be another form of wider data sharing with minimal interpretation allowing broader introspection. Workshop participants also discussed the difficulty of reporting negative data or results that support a null hypothesis. Journal articles that can accommodate these reports or databases that can host null datasets are essential for helping researchers to recognize and evaluate potential experimental avenues and use this information to understand technological and biological limits.

Reporting high-profile results to the media and general public will require communication training that not all researchers are formally taught. A good example is the Sharing Science program designed by the American Geophysical Union. The community would benefit from astrobiologists engaging in these and others like them offered by other professional societies, with the incentive of gaining better communication skills and recognition. While workshop attendees were eager for more opportunities for researchers, science communicators and journal editors to discuss the Assessment Framework in the context of a reporting protocol, it was recognized that there is also value in building better collaborations between institutional communication professionals and science journalists. These specialists can help convey a message to the public that avoids poorly verified information.

### 5.1.4 How do we encourage and support collaborative verification?

Since all participation in these collaborative efforts is voluntary, and can only be encouraged and not proscribed, community actions are the key to promote and support collaborative verification. Suggestions proposed during the workshop included:

● Dedicated sessions in scientific meetings that can be used to present the reporting protocol, discuss further improvements, establish verification support mechanisms and demonstrate the benefits of following the protocol and participating in verification activities.

● Establishing a committee or group to coordinate communication and information exchange, and otherwise support implementation of the reporting protocol. A useful step may be the *formation of an Astrobiology Society*. The ideal committee should be international, sampling the demographics of our community and the diversity of disciplines involved in astrobiology. Steps towards inclusivity could include seeking support from existing international intergovernmental organizations (e.g. the United Nations Office of Outer Space Affairs).

● Periodic assessment of the totality of evidence for life on specific targets which can be performed by the newly formed Astrobiology Society or a group constituted ad-hoc for that task.

● Creating databases (such as the LDKB) of examples of consensus standards in other fields

Verification of results may happen in several stages. Before submitting to a refereed journal, the research group may share data with another research team for a simultaneous publication of findings (discovery and possible verification). After publication, the aforementioned committee may help to build a verification strategy. A curated list of peer-reviewers for life detection claims may be helpful for editors handling journal articles related to the search for life. We recognize that some of these actions require the active participation of journal editors who are part of the scientific community but act under the policies of their journals. Editor and journal engagement is fundamental for the verification protocol. The verification process in any of the described stages would benefit from financial and organizational resources dedicated to this task.

As a group, we assume that there are ethical rules and codes of conduct in our countries and institutions that must be followed during research, publication, reviewing and collaboration that apply to the verification process. Following ethical rules and codes of conduct creates a more trusting environment within the community and favors meaningful collaborations and the verification or results. This process could facilitate building trust between scientists, the public media, and the general public and increasing public understanding of astrobiology topics. By putting in practice the protocol and verification processes as advisors and teachers we contribute to making good practices a natural way to report, collaborate and verify scientific results (Des Marais, 2008).

## 5.2 Lessons learned from other fields

Communication challenges are not unique to astrobiology. Other fields, such as climate change and pandemic-related studies, have seen an increasing spread of misinformation and disinformation in recent years. Figure 5.2 demonstrates a tool designed by the IPCC to communicate the strength of evidence for a claim as it correlates with the level of agreement in a scientific community that has evaluated the evidence. Public communication of the Assessment Framework should also strive for intuitive, informative explanations. Astrobiology can learn lessons from experts in these other high-profile, high-societal-impact fields to understand how to disseminate clear information and best incentivize a proper balance between claims and caveats in publications.

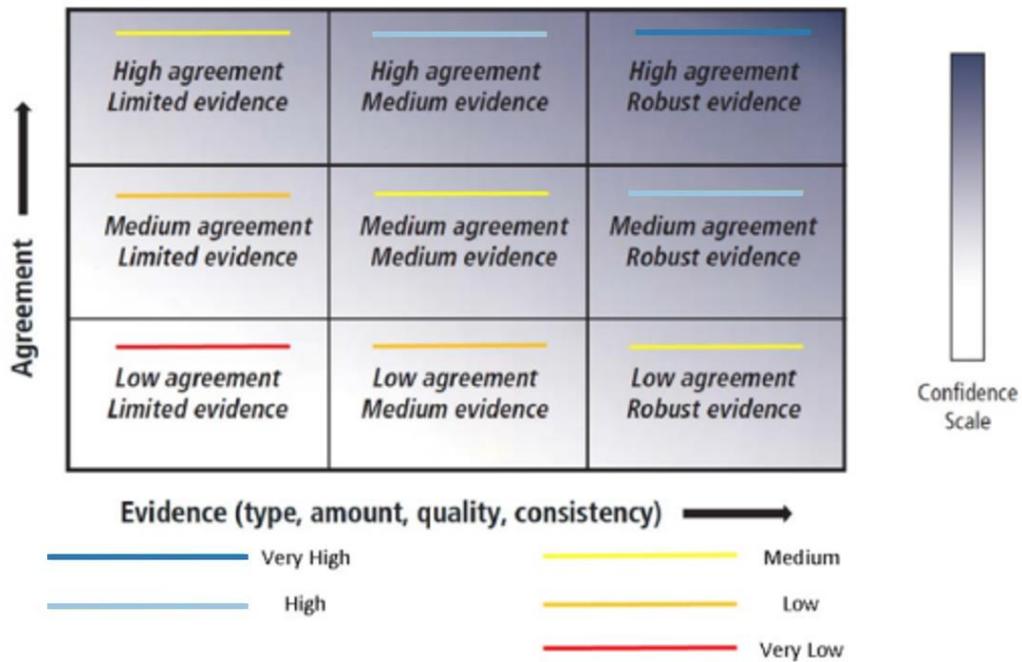

Figure 5.2 Confidence levels are a combination of levels of agreement and evidence. There are five levels shown with colors (after IPCC 2013).

## 6:  Findings and Future High Priority Research Avenues

A discussion of the best practices needed to confirm the detection of life elsewhere in the Universe is critically needed, and the Biosignatures Standards of Evidence Workshop was able to bring together astrobiologists from early Earth, Mars, icy world and exoplanet life detection fields to address this common need.  It was agreed that it is particularly important to accurately portray the search for life as a multi-step, potentially multi-mission process, rather than one achievable with a single investigation (e.g., Green et al., 2021). Forging a stepwise biosignature Assessment Framework is in the best interest of science and scientists, and can allow us to provide better communication of the significance of new discoveries to the general public. This framework allows each new report or planned mission to be evaluated for its contribution to the larger goal, and with the realization that even if life is *not* detected, the field is progressing.

In the following subsections we summarize key findings and future avenues of research identified by the community, to continue the discussion of how best to assess and communicate claims of life detection.

## 6.1 Future Work to Enhance the Flexibility and Power of the Assessment Framework

It has been stressed previously (Green et al., 2021) that there is a continuum from a low to higher certainty level in whether or not a given phenomenon is due to life, and that any additional data can help further our collective understanding. Trust in the veracity of a biosignature claim has two components: (1) uncertainty in the detection due to measurement error and potential contamination and (2) uncertainty in the interpretation due to lack of understanding of environmental context, and abiotic and life processes.  Uncertainty in the detection and identification component is more readily quantifiable, and likely requires only field- or even measurement-specific expertise.   Confidence in the interpretation, although harder to quantify, could be increased via improved understanding of environmental context, the likelihood of false negatives and false positives, and an improved understanding of the mechanisms of biosignature generation.   The interpretation therefore involves a broader expertise base and more members of the community.  While workshop participants adopted the generalized steps included in the framework, the measurement constraints and interpretation needs of different astrobiological fields let to the realization that the framework does not need to be linear and may be more effective if it can be iterative (see Section 2). However, to reduce the overall burden on the astrobiology community, it was considered critical to undertake a rigorous study of the signal detection and identification prior to the necessarily interdisciplinary assessment of whether or not the signal was due to life, as the latter undertaking involves a significantly larger number of scientists.

During discussions at the workshop, participants across disciplinary divides agreed on the importance of:
- multiple lines of evidence for both detection and interpretation
- exhaustive examination of contamination and other sources of false positives
- discriminating potential biosignatures from abiotic and prebiotic chemistry
- geological context/habitability
- biosignature preservation/atmospheric lifetime potential

They also identified topics that are likely to require significant additional community discussion to resolve, including:
- further development of the framework to generate a more detailed set of worked examples that can serve as field-specific guidelines for biosignature assessment
- enhancement of the framework to incorporate statistical methodology at each step, and the development of a ranking or numerical scheme for certainty in biosignature detection and interpretation.

When claiming the detection of an extraterrestrial biosignature, multiple lines of evidence are crucial, in both corroborating a detection, and in ruling out potential contamination.  The strongest lines of evidence for life would be multiple measurements of a given biosignature, or even a suite of biosignatures (NASEM, 2019). Previous work has identified multiple different types of biosignatures including: organic molecules, minerals, macro- and micro-structure, chemistry and isotopes (Mustard et al., 2013). Therefore, multiple measurements, and identification of multiple potential biosignatures, would provide the strongest evidence of life. Biosignature assessment can be strengthened by forming and testing hypotheses e.g "If the detected signal(s) are due to life, we should also see phenomena x and y; do we see x and y?"

Examination of the planetary context and the potential habitability of an environment (see Questions 3 and 4) would help to rule out abiotic processes and strengthen the argument for detection of life.  Community work to survey, identify and understand likely abiotic processes that can mimic life's impact on its environment is a critical preparatory step for applying the Assessment Framework. Participants agreed on the need to undertake laboratory work to better understand abiotic expressions without the imprint of the billions of years of biotic influence we see on Earth, although abiotic processes can be identified and studied via a variety of means.  Once identified and characterized, predictions can be made for measurable discriminants that would identify the abiotic process.  For example, Question 3 of the Assessment Framework would require researchers to also invert the previous idea and ask "if this signal *isn't* due to life, but is instead due to the following abiotic process/characteristic, then we should see phenomena p and q"; do we see p and q? This covers concerns vis-a-vis abiotic biomorphs such as the "nanobacteria" observed in meteorite ALH84001 or other systems (McKay et al., 1996; Schieber & Arnott, 2003). Therefore, the tests of the impacts of life on the environment can also be addressed here.

The ability to interpret whether or not a signal is due to life can also be strengthened using a number of strategies including prior modeling to understand the potential for abiotic mimics in a given planetary environment, and the use of data from analog environment investigations and laboratory simulation experiments (Assessment Framework Questions 3 and 5; e.g., Hays et al., 2017 and Mateo-Martí et al., 2010). Although no analog or experiment is perfect, the data from these studies can help us understand potential discriminants between biological and false positive processes.  False negatives can be somewhat negated by identifying and, if possible, avoiding environments and samples in which they are more likely to occur, and in using larger samples to increase the probability of detecting sparse life.  Finally, discriminating between life and non-life hypotheses will be supported by continued development of extensive libraries of known relevant abiotic and biotic signatures for both extant and extinct life (e.g., The Life Detection Knowledge Base; Chan et al., 2019) and better defining the limits of life to use as standards.

In addressing Question 4 and 5, on the likelihood that life would produce a signal in this environment,  detection of a biosignature in an environment that is known to be habitable for terrestrial-like life would provide a line of evidence that would be easier to interpret as supporting the life hypothesis. However, potential detections of life should not be limited to environments that are known to be habitable for terrestrial life, and should also include the search for "life as we don't know it", potentially through agnostic biosignatures.

Biosignature preservation potential was identified as another central theme that potentially impacts target selection and multiple Questions in the Framework, and that may require additional work to explore its ramifications.  Preservation in this sense can refer to a broad range of processes that span disciplines and measurement techniques, from preservation over time in the rock record, to the chemical lifetime and persistence of biosignature gases in a planetary atmosphere.

The participants also discussed how the framework could potentially guide different stages of mission development, from concept to operation.  The framework could be used to help develop a mission concept's top-level science goals and instrument suite, and the specific investigations which could meet those goals. It may also help guide and ensure the inclusion of measurement capabilities for verification, contamination identification, and environmental characterization, which would be needed to confirm biosignature detection, and help with biosignature interpretation.   Additionally, if we as a community assign a high priority to the return of space samples for biosignature investigations on Earth, this will necessitate the development of additional protocols for sample storage, handling, transport, sharing, and curation in the context of international partnerships (Meyer et al., 2020).

Given that the workshop included only 3 half days of discussion, several important topics were identified for further discussion in subsequent workshops or other fora.  Two of the highest priorities include further work to develop more detailed "worked examples" for specific biosignatures and measurement techniques, and significant further work needed to incorporate statistical methodology into methods to address each Question in the Framework. The workshop participants were able to agree on a fundamental, generalized framework for biosignature assessment, guided by a series of specific questions that could be interpreted for applicability and applied to a broad range of life detection claims.  Initial "Worked examples" of how this framework might be applied to a few different biosignatures and measurement techniques are provided in Section 3.  However, specific fields may wish to pre-identify and further develop best practices and field-agreed-upon actions for a particular biosignature and measurement technique, or otherwise discuss more detailed treatments that may require further community input to develop and adopt.

Another area where discussions were initiated but not finished at the workshop included evolving the Framework to incorporate statistical methodology in answering each Question. There was strong agreement that a systematic understanding of additive uncertainty should be part of any evidence validation activity. Further, workshop participants expressed the need for research that addresses the data and meta-data needs in order to apply statistical methods such as a Bayesian analysis. The value of a Bayesian methodology has previously been endorsed in the Astrobiology Strategy report (NASEM, 2018) since this method incorporates assumptions and prior-information and can then constrain probability of life even without a positive detection (Walker et al., 2018; Catling et al., 2018). Incorporating statistical methodology into the Framework represents a significant amount of future work, but it would help researchers include statistical needs into data collection processes early in a life detection effort.  Part of the statistical discussion may also incorporate the value of machine and deep learning (ML/DL) methods are also important to distinguish signal from noise when data is not straightforward, or when there are multiple data sources. Additionally, a detection implies a probabilistic approach which can include models (i.e., Bayesian) that can be easily adapted to the multiple lines of evidence and include updates on the evidence as well.  Developing mechanisms to encourage collaboration between biosignature scientists and statistical specialists was identified as a particularly fruitful avenue for enhancing the rigor of biosignature assessment.

## 6.2 Expanding our Ability to Detect and Identify Agnostic Biosignatures

In addition to expanding our knowledge of potential biosignature targets for known metabolisms, potential detections of life as we don't know it would strongly benefit from the contributions of "agnostic biosignatures" (biosignatures that do not rely on life on Earth as an analog or target analyte). An improved understanding of agnostic biosignatures could greatly expand the search for life beyond current searches that include more Earth-centric assumptions about metabolism, ecology, and physiology. An agnostic approach could include applications from information theory and computational simulations (Keefe et al., 2010; Goodwin et al., 2015; Cronin and Walker 2016; Cabrol, 2016) as well as more fundamental qualities like environmental complexity and broader physical and chemical impacts due to life. Although agnostic biosignature interpretation requires a thorough knowledge of environmental context, it can be less biased due to fewer assumptions about context and biological markers, while preserving some of the fundamental laws of nature (i.e., evolution - evolutionary algorithms; entropy - both physical and informational; network dynamics, scalability, etc.). Lack of knowledge of potential agnostic biosignatures could also increase the chance of a false negative, where besides false negatives from worlds with sparse life and weak signals, we could miss "life in its most general form".

Other considerations discussed included the possibility that chemistry that appears auto-catalytic may not be due to life, but instead may be a form of prebiotic chemistry. An improved understanding of the continuum of chemistry from non-life to life would also help identify possible biosignature false positives. If evolution of biology from abiotic components is a continuum, we may observe some intermediate stage, in which case knowledge of the prebiotic/biotic boundary will greatly enhance our ability to identify whether the environment contains life.

As a final point, discussion highlighted the need to make assumptions regarding life detection clear and explicit. For example, the current working hypothesis for life detections is the *null hypothesis* - that life is not the explanation for the detection, and this assumption should be made more explicit. Other potential assumptions or definitions that may produce a lack of clarity include how habitability is defined in a given context, whether is it assumed that habitability refers to the conditions required to support "known life", rather than life as we don't know it, or the requirement for surface liquid water for exoplanet habitability. These can be important arguments strengthening the detection of life, but can nonetheless represent assumptions that should be explicitly stated.

## 6.3 Community Engagement Structures

While the main workshop goal was the development and adoption of a generalized biosignature Assessment Framework, workshop participants held a brainstorming session on the last day to generate topics that would be continued in a future workshop focused on designing "best practice" reporting protocols that guide biosignature discovery public announcements. Workshop participants considered and analyzed the processes and constraints associated with contemporary scientific discovery, what discussed ways that scientists, journalists, and others involved in science communication could potentially compensate for structural challenges. (Figure 6.1)

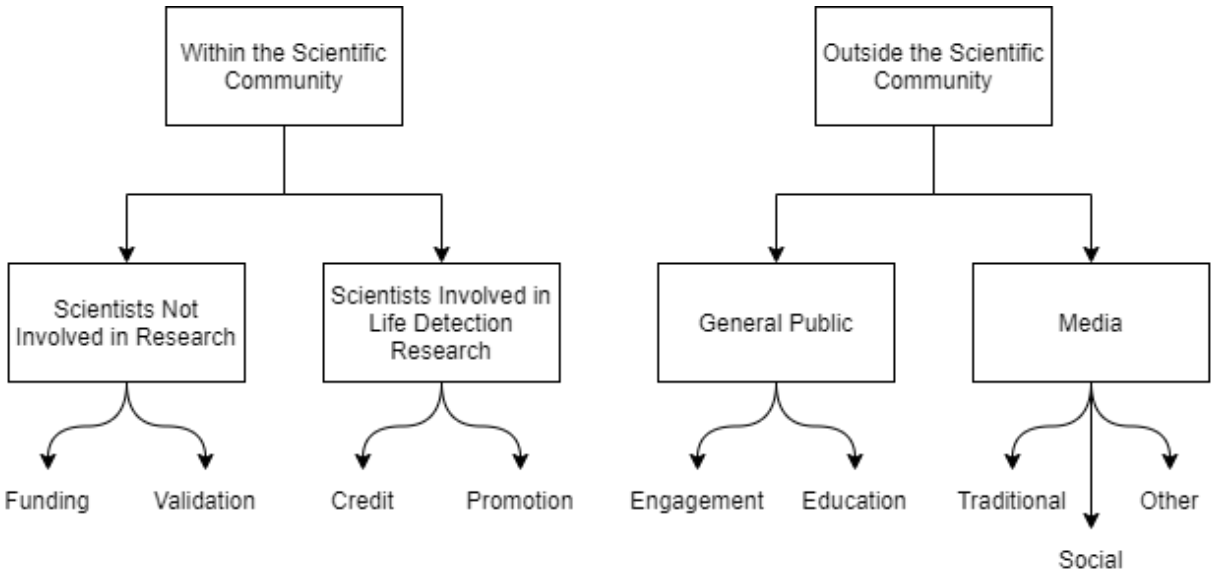

***Figure 6.1*** *Summary of Stakeholders Involved in Life Detection Findings Dissemination. There are numerous communities, including potentially new ones, that will engage in the scientific discussion of the detection of life elsewhere in the Universe. The squares in this diagram represent stakeholder communities, and the bottom row identifies principal incentives for these communities to be interested in claims of life detection. Scientists are motivated by the possibility of scientific credit, promotion, funding of follow-up work and scientific validation. Outside the scientific community, there is demand for both the engagement and education of the general public as moderated by the media.*

As a community, workshop participants came to a few points of consensus on supporting aspects of protocols for biosignature assessment and reporting, some of which may take significant effort to slowly change the status quo. These avenues include:

- Encouraging funding support specifically for verification work by independent groups, either pre- or post- publication, including the possibility of pursuing international options for a cohesive global support network
- Expanding content/access of open servers that host data, pre- and post-publication and encouraging voluntary sharing of data and results.
- Working with publishers to support simultaneous publishing and promotion of discovery and verification papers, stepwise reporting of results including caveats, and reporting of null results.
- Similarly modifying our scientific culture and workplace incentives to also value rigorous, iterative science, validation and null results, in addition to high-profile discoveries.
- Forming an independent international committee to help coordinate information exchange and verification efforts, which could be a new Astrobiology Society.
- Periodic review and assessment of the totality of evidence for life on specific targets as an Astrobiology Society function or via an international committee.
- Outreach to scientists, journal editors and journalists to present the finally agreed upon reporting protocol and describe its potential benefits.

- Review and learn from other high-profile research areas such as climate change, LHC/CERN, public health (i.e., vaccinations), epidemiology, etc.

Any and all of these steps could be impactful in increasing transparency, and producing the most robust, peer-reviewed science while combating "disinformation"—as messaging can slip from the control of scientists via the rapid dissemination that comes with a digitally connected population. In particular, incentivizing scientists to collaborate on fewer, more robust, and so higher impact papers could come from the institutions they work at (e.g., universities, federal laboratories, etc.) as well as funding agencies (e.g., NASA, ESA, etc.). If each author/participant involved benefited in some concrete way from the publication(s), then this would in turn incentivize more collaborative, and potentially more rigorous science. However, a mechanism would be needed to distribute the credit for the finding to not only the first author and one particular funding agency, but also the other authors and potentially outside groups validating data.

**Funding Mechanisms to Facilitate Collaboration**
If we take other constraints into consideration, funding additional, potentially rapid response work to validate (or not) a biosignature, either via alternate analytical methods or by reproducing the results using identical methods used by the original group, would be extremely impactful in both confirming the research as well as building trust with the general public. Funding is an obvious mechanis to transform how our research is done, and the public communication of that research. Examples of a potentially effective mechanism include the short term rapid response funds managed by agencies such as the NSF and the International Ocean Drilling Program that can direct funds towards immediate needs (e.g., in the event of a geohazard). A fund of this kind could be activated towards validation needs in the event of a promising life detection claim.

Several examples also exist of beneficial public-private partnerships for scientific research. In the United States the National Institutes of Health (NIH), and other key governmental agencies such as the FDA and NSF have foundations associated with them (secondary funding support mechanisms) that leverage the agency's peer review expertise to identify key research. This, coupled with the agencies' funding management process is used to distribute private funding to targeted individual research as well as to topical research foci. A "Foundation for Astrobiology" with a specific focus on life detection verification could provide an opportunity for funding from philanthropists to be managed by a government agency and a public board of directors.

There are other examples of national and international collaborative efforts in a wide range of disciplines, including the establishment of professional societies that might serve as guides for the Astrobiology community, and could be further explored. Existing international collaboration with astrobiology-focused funding responsibilities shared across borders include a number of astrobiology training courses around the world. These include the Iberoamerican School of Astrobiology and the International School of Astrobiology. These are funded by NASA and international governmental funding, including UNESCO. Notably, the former courses are taught in Spanish, thus being deeply inclusive to a significant source of human capital while training a new generation of students in astrobiology. Given the inherently interdisciplinary nature of the search for extraterrestrial life, enlisting the intellectual and collective assistance of scientific communities around the world may well turn out to be critical for the acceptance of a novel finding. Participants also expressed interest in the formation of an astrobiology professional society that as part of its activities could identify an independent body to coordinate and support life detection verification and reporting protocol activities, and associated resources.

**Incentivizing The Assessment Framework**

While workshop participants agreed upon the need to devise a framework for biosignature assessment and reporting, participation would necessarily be voluntary, and a much more difficult conversation involves how we might incentivize adoption of this framework. We would argue that this framework, if implemented correctly, could be generally applicable to any number of high profile scientific endeavors that will attract a great deal of attention from the scientific community as well as the media and general public. But how can we incentivize the adaptation of existing scientific practices towards a more collective, collaborative response? We wrestle with this question below, and suggest a number of ways that the scientists doing this work, as well as the larger scientific community, could be strongly encouraged to both utilize the framework *and* abide by a principled public discourse. The overarching motivator comes down to funding.

There are far more reasons for scientists NOT to use this framework or follow the outlined process for a reputable reporting protocol without some additional incentives. Many, if not all, come down to funding and recognition. In many ways, funding is what drives science in the 21st century, and therefore recognition is critical for continued engagement of funding. Programs to incentivize and reward biosignature validation could follow examples in medicine and social sciences that fund reproducibility studies. In contrast, private entities have tried to incentivize scientists through high-profile targeted awards focused on risky, "moonshot" research and conspicuous technology demonstrations. Even with the promise of personal financial gain and fame, it is not clear that these competitions lead to a robust advancement in science. Transparent, peer-reviewed funding programs have greater promise to provide the necessary background to bolster the Assessment Framework.

Committed funding for both the research and the validation research (Fig. 5.1), would incentivize following the framework itself, including a reasonable approach to reporting protocols. If as a community, we remove the potential threat of detriment for the original authors (and therefore, a more equally distributed risk to all the authors) this would promote a more rational approach to disseminating the findings. Distributing the reward or recognition of such a groundbreaking discovery would also be more in line with research on climate change (e.g., the IPCC) and work produced by the Large Hadron Collider (i.e., CERN). This diffusion of responsibility and recognition could create a process where each participating scientist (either original research or validation research) shares credit, as well as taking ownership, of their individual contribution, and promote engagement in collaborative research.

**6.4 Additional ideas from the community in response to the workshop report draft**

The above document describes the discussions and outcomes from the community workshop, but during our public comment session, we also received input from the broader community, including those who did not attend the workshop.  Many of these suggestions endorsed or strengthened avenues for future work identified at the workshop, while others proposed new efforts.  Additional community suggestions included:

- Prioritization of research to develop a theory of life, that could support biosignature detection.
- The integration of biosignatures and technosignatures.  Parallel efforts are underway in these two communities.  In particular "chemical" or "climate" technosignatures could represent a sign of advanced life and may need to follow a similar assessment framework, including identification of potential false positives. Similarly protocols for distinguishing radio SETI signals from terrestrial contamination (Sheikh et al., 2021) have parallels with the biosignature assessment framework.
- Leveraging Earth analog environments and laboratory studies to help discriminate abiotic environmental characteristics from life's impact on that environment (e.g. Martins et al., 2017)

## Acknowledgements

This fully-virtual, pandemic-embedded workshop would not have been as effective, or indeed probably even possible, without the  invaluable support of the Know  Innovation team led by Andy Burnett. We thank Andy and the Know Innovation for their support in organizing and facilitating broad community interaction at the workshop, including multiple forms of virtual interaction across several time zones!  We would also like to thank Liza Young for her outstanding editorial and graphic support.

## Glossary of Acronyms

ALMA: Atacama Large Millimeter/submillimeter Array

CERN: Conseil Européen pour la Recherche Nucléaire , or European Council for Nuclear Research

CIA: Collision-Induced Absorption

CMOLD: Complex Molecules Detector

CNRS: Centre national de la recherche scientifique (French National Centre for Scientific Research)

COVID-19: Coronavirus Disease 2019

DAVINCI: Deep Atmosphere Venus Investigation of Noble gases, Chemistry, and Imaging

DL: Deep Learning

DLR: Deutsches Zentrum für Luft- und Raumfahrt (German Aerospace Center)

ESA: European Space Agency

ETH: Eidgenössische Technische Hochschule

ETI: Extraterrestrial Intelligence

ExoPAG: Exoplanet Exploration Program Analysis Group

FDA: U.S. Food and Drug Administration

GCMS or GC-MS: Gas Chromatograph Mass Spectrometer

GISS: Goddard Institute for Space Studies

GSFC: Goddard Space Flight Center

HabEx: Habitable Exoplanet Observatory

HITRAN: High Resolution Transmission

HZ: Habitable Zone

InSAR: Interferometric Synthetic Aperture Radar

IPCC: Intergovernmental Panel on Climate Change

ISO: International Organization for Standardization

JAXA: Japan Aerospace Exploration Agency

JPL: Jet Propulsion Laboratory

JWST: James Webb Space Telescope

LATMOS: Laboratoire atmosphères, milieux, observations spatiales

LCMS or LC-MS: Liquid Chromatography–Mass Spectrometry

LCROSS: Lunar CRater Observation and Sensing Satellite

LDKB – Life Detection Knowledge Base

LHC: Large Hadron Collider

LIFE: Large Interferometer For Exoplanets LPI: Lunar and Planetary Institute

LLC: Limited Liability Company

LR: Labeled Release

LSW: Landessternwarte Königstuhl

LUVOIR: Large Ultraviolet Optical Infrared Surveyor

MEPAG: Mars Exploration Program Analysis Group

MISS – Microbially-Induced Sedimentary Structures

ML: Machine Learning

MTBSTFA: N-Methyl-N- (Tert-Butyldimethylsilyl)trifluoroacetamide

NAO: National Astronomical Observatory

NASA: National Aeronautics and Space Administration

NASEM: National Academies of Sciences, Engineering, and Medicine

NExSS: Nexus for Exoplanet System Science

NfoLD: Network for Life Detection

NIH: National Institutes of Health

NMR: Nuclear Magnetic Resonance

NOW: Network for Ocean Worlds

NSF: National Science Foundation

OPAG: Outer Planets Assessment Group

ORAU: Oak Ridge Associated Universities

SAM: Sample Analysis at Mars

SETI: Search for Extraterrestrial Intelligence

SNR – Signal-To-Noise Ratio

STM: Science Traceability Matrix

TDEM: Technology Development for Exoplanet Missions

TRAPPIST: Transiting Planets and Planetesimals Small Telescope

UC: University of California

UCLA: University of California, Los Angeles

UNESCO: United Nations Educational, Scientific and Cultural Organization

UPR: Universidad de Puerto Rico

US: United States

UV-VIS-NIR: Ultraviolet-Visible-Near-Infrared

VERITAS: Venus Emissivity, Radio science, InSAR, Topography, And Spectroscopy

VRE: Vegetation Red Edge